\documentclass[pre,aps,superscriptaddress,twocolumn]{revtex4-1}   	

\usepackage{amsmath,bm}
\usepackage{graphicx}
\usepackage{subfigure}
\usepackage{soul}
\usepackage{bbold}
\usepackage{color}

\usepackage{comment}

\newcommand{\beq}{\begin{equation}}
\newcommand{\eeq}{\end{equation}}

\newcommand{\p}{\partial}
\graphicspath{{figures/}}

\DeclareMathOperator{\Tr}{Tr}

\begin{document}

\title{Dynamics of undulatory fluctuations of semiflexible filaments in a network}

\author{Jonathan Kernes} 
\affiliation{Department of Physics and Astronomy, UCLA, Los Angeles California 90095-1596, USA}

\author{Alex J. Levine}
\affiliation{Department of Physics and Astronomy, UCLA, Los Angeles California 90095-1596, USA}
\affiliation{Department of Chemistry and Biochemistry, UCLA, Los Angeles California 90095-1596, USA}
\affiliation{Department of Computational Medicine, UCLA, Los Angeles California 90095-1596, USA}

\date{\today}

\begin{abstract}
We study the dynamics of a single semiflexible filament coupled to a Hookean spring at its boundary. The 
spring produces a fluctuating tensile force on the filament, whose value depends on the filament's instantaneous end-to-end 
length. The spring thereby introduces a nonlinearity, which mixes the undulatory normal modes of the filament and changes their dynamics. 
We study these dynamics using the Martin-Siggia-Rose-Janssen-de-Domincis formalism, and compute the time-dependent 
correlation functions of transverse undulations and of the filament's end-to-end distance.  The relaxational dynamics of the modes 
below a characteristic wavelength $\sqrt{\kappa/\tau_R}$, set by the filament's bending modulus $\kappa$ and 
spring-renormalized tension $\tau_{R}$, are changed by the boundary spring.  This occurs near the cross-over frequency 
between tension- and bending-dominated modes of the system.  
The boundary spring can be used to represent the linear elastic compliance of the rest of the filament network to 
which the filament is cross-linked.  As a result, we predict that this nonlinear effect will be observable in the dynamical 
correlations of constituent filaments of networks and in the networks' collective shear response.  The system's 
dynamic shear modulus is predicted to exhibit the well-known crossover with increasing frequency from $\omega^{1/2}$ to $\omega^{3/4}$, 
but the inclusion of the the network's compliance in the analysis of the individual filament dynamics 
shifts this transition to a higher frequency. 
\end{abstract}
\maketitle

\section{Introduction}

Semiflexible filaments networks underlie the structure of a number of biological materials, including the cytoskeleton 
and the extracellular matrix of tissues~\cite{Pritchard2014,Broedersz2014,Chen2010}. The mechanical properties of such materials depend on the mechanics of their individual filaments. 
These semiflexible filaments are essentially inextensible, with lengths less than their thermal persistence length, indicating a large 
bending rigidity $\kappa$ that keeps them oriented along a mean direction.

Filamentous networks exhibit a number of interesting mechanical properties that differ from typical elastic continua, such as 
nonaffine deformation~\cite{heussinger2007, Fernandez2009} and negative normal stress~\cite{Janmey2007, Kang2009}.  There is 
now a well-developed theory connecting the
tension response of individual filaments to the linear collective shear response of their networks: $G(\omega)$.  Due to the 
appearance of multiple time scales in the networks' dynamics, 
$G(\omega)$ exhibits a rich variety of behaviors~\cite{Gittes1997,gittes1998G,granek1997R,morse1998visc}.

\begin{figure}
\includegraphics[scale=0.5]{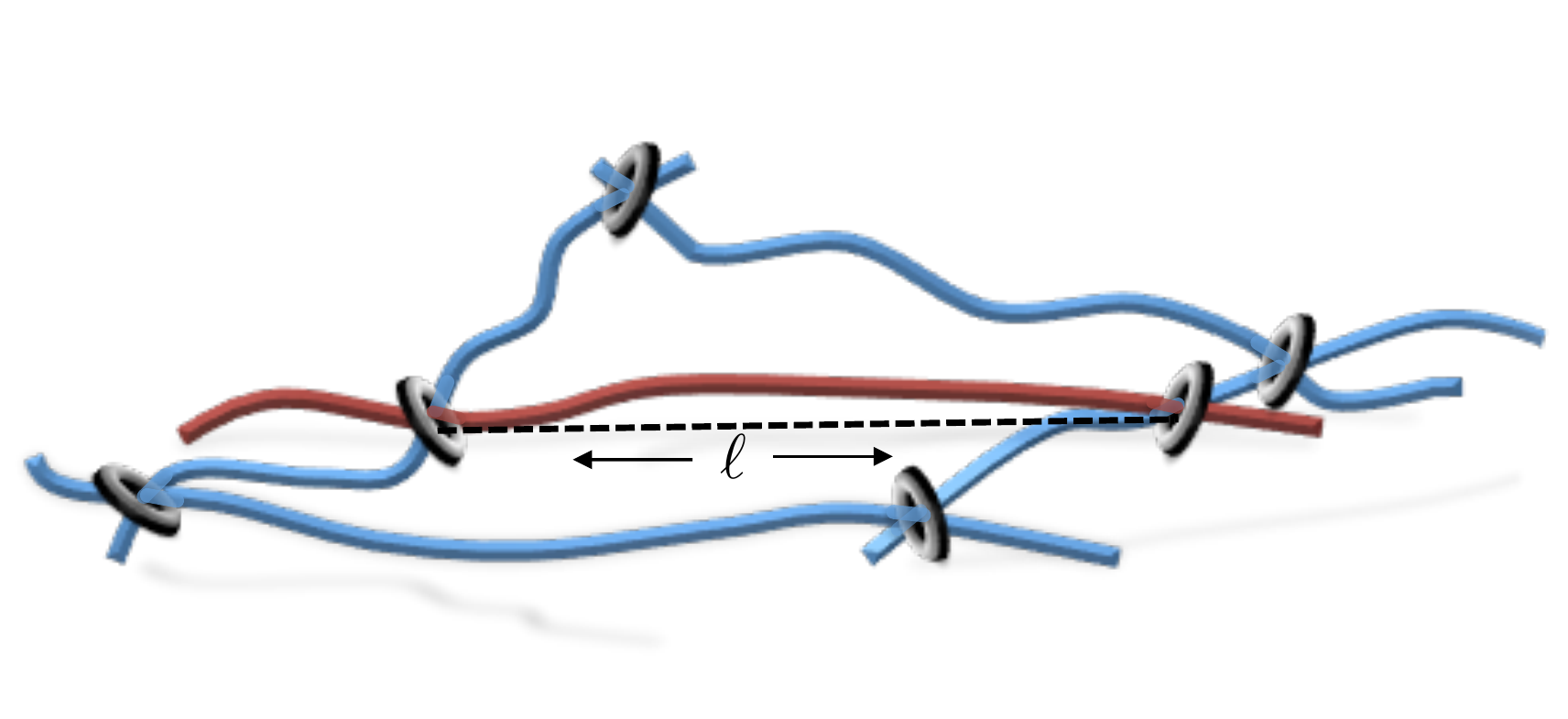}
\includegraphics[scale=0.76]{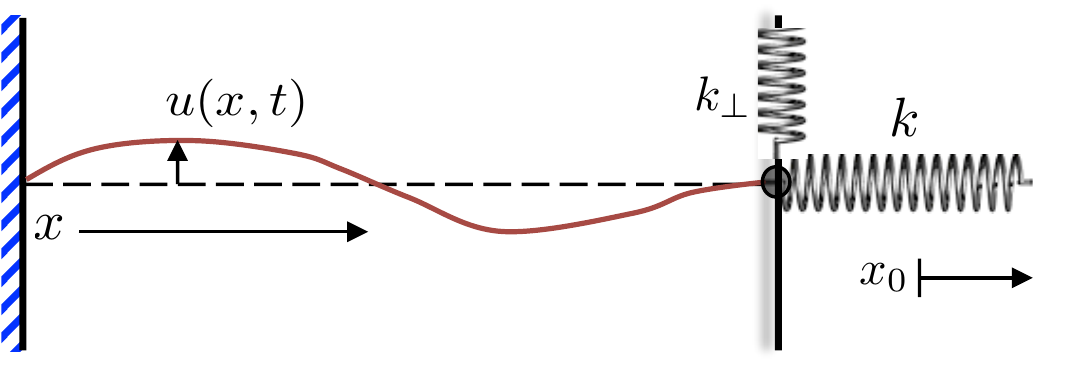}
\caption{(color online)  Top: visualization of a particular filament (red) cross linked into a network of similar filaments (blue). The cross links are 
represented by black and gray rings. Bottom: schematic diagram of a single semiflexible filament. The left endpoint is pinned, 
and the right attached to a longitudinal spring with spring constant $k$
and a transverse spring with spring constant $k_{\perp}$. These represent the elastic compliance of the network. We focus on the 
effect of the longitudinal spring. Both endpoints are subject to torque-free boundary conditions.}
\label{fig: schematic}
\end{figure}

There is currently considerable interest in local microrheological probes of tension within the network at the single filament scale. Individual filaments in 
network are subject to thermal fluctuations. Their fluctuation spectrum is, in part, controlled by the filament's 
mechanical boundary conditions imposed by its coupling to the 
rest of the network.  For example, the fluctuations of the (red) filament in Fig.~\ref{fig: schematic} are modified by that filament's mechanical 
coupling via cross links (black/gray circles) to the surrounding network of 
(blue) filaments.  These boundary conditions include the tension imposed on the filament, allowing, in principle, 
one to extract local tensions from the observations of the stochastic
undulations of individual strands within the network. The technique is called {\em activity microscopy}~\cite{Lissek2018,kernes2020equilibrium}.  

In our previous paper~\cite{kernes2020equilibrium}, we examined how the surrounding network, 
including its elastic compliance and state of tension, affects the equilibrium fluctuation spectrum
of the transverse undulations of a constituent filament in the network.  In this manuscript, we expand our 
analysis to dynamics, looking at the time-dependent correlation and response 
functions of both individual transverse modes of the filament and its end-to-end distance.  These results will 
be important for future work on the frequency-dependent 
{\em nonequilibrium} fluctuations of network filament segments driven by 
endogenous molecular motors~\cite{mizuno2007nonequilibrium,mackintosh2008nonequilibrium,mizuno2008active,levine2009mechanics}.

We model the mechanical boundary conditions on the semiflexible filament by both a mean state of tension $\tau$ 
and a linear elastic compliance, representing the surrounding network.  This 
elastic compliance may be thought of as attaching the filament's end to a pair of Hookean springs, one longitudinal spring 
aligned with the mean extension of the filament and one perpendicular to it.   These
springs have spring constants $k$ and $k_{\perp}$ respectively.  
The lower panel of Fig.~\ref{fig: schematic} shows a schematic illustration of the system, whose dynamics will be the focus of this manuscript. The perpendicular spring changes the spatial structure of the eigenmodes of filament deformation -- see Appendix~\ref{app: transverse spring}. The longitudinal spring is the most interesting, as it introduces a nonlinearity into the filament's Hamiltonian even in the limit of 
small bending.  The origin of this nonlinearity (explained more fully below and in Ref.~\cite{kernes2020equilibrium}) is that the state of tension in the filament depends on the instantaneous
projected length of the filament.  
Since we work in the limit that the filament is inextensible, tension propagation is instantaneous, and the longitudinal spring introduces a term that is non-diagonal in the Fourier modes of the filament's undulations, but local in time. As a result of this nonlinearity, the equilibrium
fluctuation spectrum~\cite{kernes2020equilibrium} of the filament and its dynamics are controlled by the set of parameters $(\kappa, \tau, k)$.

The nonlinearity introduced by the boundary compliance (due to the rest of the network) 
alters the power spectrum of the variations of the projected length 
of a filament's end-to-end distance when it is cross linked into a network. We study that here. 
Using our model, we  also calculate the response of that distance to applied forces. The time-dependent, single-filament response can be then be used to 
calculate the dynamic shear modulus and compliance of the network by well-known methods~\cite{gittes1998G}. The most direct 
experimental test of our theory, however, is to be found at the 
single filament level.  We propose that one can directly measure the relaxational dynamics of a single filament anchored to a 
substrate and attached to a bead held in an optical trap~\cite{Starrs2002,Addas2004,Latinovic2010}.  
In such a configuration, the trap provides a longitudinal spring of known (in principle) spring constant. By 
moving the trap's center, one can measure the changes in filament's 
fluctuations as a function of tension. In addition to passive measurements, one should also be able to 
actively measure the response function of the filament's end-to-end distance by 
driving it via the sinusoidal oscillations of the trap's center. We predict that the new effect 
associated with the elastic compliance of the trap will be most evident at small values of applied tension.

The remainder of this manuscript is organized as follows.  We introduce the model, including the stochastic equation of 
motion (Langevin equation) of the filament using slender body dynamics, in Sec.~\ref{sec: The Model}. 
Due to the nonlinearity in the Hamiltonian, we cannot solve these dynamical 
equations exactly. Instead, in Sec.~\ref{sec: perturbation theory} we first compute the linear response to transverse 
applied loads in the wavenumber domain to second order in the longitudinal spring constant.   
From this, we determine the dynamical two-point function
$\langle | u_p(\omega)|^2 \rangle$ in Fourier space. These perturbative 
calculations are organized using the Martin-Siggia-Rose/Janssen-de-Domincis (MSRJD) 
functional integral formalism~\cite{tauber2014critical}.   Within this diagrammatic expansion, 
we comment on various classes of diagrams and propose an approximation using a resummation of the dominant terms of the perturbation series.  

To make these approximations precise, we develop an effective field theory (in Sec.~\ref{sec: auxiliary field theory}), whose mean-field solution 
reproduces the resummation of the dominant terms. The mean-field solution is a type of dynamical self-consistent theory, which we analyze in Sec.~\ref{sec: MFT}. 
This self-consistent approach allows us to explore non-equilibrium dynamics and the time-dependent response of pulled filaments. In Sec.~\ref{sec: fluctuations} 
we expand our effective field theory about its mean-field solution, allowing us to compute fluctuations, particularly of normal modes 
and the filament's projected length. These corrections are found to be rooted diagrammatically in the random phase 
approximation borrowed from solid state physics. By considering the terms arising at higher orders in the expansion 
about the mean field, we identify the various classes of diagrams postulated from the second-order perturbative 
result, thus determining the validity of our initial approximations. We conclude with a discussion of our 
results in Sec.~\ref{sec: conclusion}, where we discuss the expected 
experimental signature of the filament's mechanical boundary conditions on its dynamics. 
The reader interested primarily in those predictions is encouraged to turn first to that section.  

We find two principal effects of the longitudinal spring. The primary one is a renormalization of the tension by the 
mean force of the spring, which can be schematically viewed as $\tau \to \tau + k \langle \Delta \ell \rangle$. Even if 
one tunes the applied tension to a small value,  the spring, responding to the fluctuations of the end-to-end 
filament distance, will impose a tension on its own.  The spring thus adds an additional energy scale that competes with the 
work done by the imposed tension. For small $\tau$, we can approximate this by using the spring-free 
result $\langle \Delta \ell^\text{free} \rangle = k_\text{B} T \ell^2/12 \kappa$~\cite{mackintosh1995}. 
The longitudinal spring constant becomes significant when $k$ reaches at least $k=k^* \approx 12 \kappa \tau/k_\text{B}T \ell^2$.  
Secondly, the nonlinearity generically reduces the effective longitudinal spring constant $k$, as a result 
of the nonlinearity transferring the elastic energy amongst the normal modes of filament 
deformation to a more energetically favorable configuration. This effect is primarily seen in the dynamical 
projected length fluctuations. At high frequencies, this effect goes away 
so that the bare spring constant once again becomes observable.

\section{Filament dynamics}
\subsection{The Model}
\label{sec: The Model}

The filament Hamiltonian with the spring-induced nonlinearity was discussed earlier~\cite{kernes2020equilibrium}, but 
we briefly reintroduce it here. Since the filament of length 
$\ell$ is assumed to be nearly straight, we work in a Monge representation, omitting overhangs, so we 
may specify the filament's configuration by its transverse coordinate $u(x)$ at a 
distance $x$ along the mean orientation. Here we work in two dimensions, with the understanding 
that in three dimensions the dynamics simply involves two copies of the fluctuations 
considered here, one for each polarization state of the undulations.  Where necessary, we  later 
mention the inclusion of both transverse degrees of freedom. 
We treat the filament as being inextensible.  Tension propagation is instantaneous. The 
change in the filament's projected length due to bending is 
given to quadratic order in the transverse displacement by 
\beq
\label{eq:length}
\Delta \ell = \frac{1}{2} \int_0^\ell \left( \partial_{x} u\right)^2 dx. 
\eeq

The Hamiltonian of the filament with bending rigidity $\kappa$, under tension $\tau$,
and coupled to a longitudinal spring with spring constant $k$ is
\beq
\label{eq: H}
H = \frac{\kappa}{2} \int_0^\ell dx  \left(\partial_{x}^2 u \right)^2 + \tau \Delta \ell + \frac{1}{2}k \Delta \ell^2,
\eeq
where $\Delta \ell$ is the amount of the length of the filament taken up by its undulations -- see Eq.~\ref{eq:length}. 
For notational convenience, the spring constant $k$ 
used here is equal to $4 k$ in Ref.~\cite{kernes2020equilibrium}. 

The tension $\tau = \tau_\text{applied} + k x_0$ can be freely adjusted using the externally applied tension  
$\tau_\text{applied}$, or by adjusting the anchoring point of the longitudinal 
spring $x_{0}$.  By a suitable choice of $x_{0}$, it can be made to vanish. We assume that 
filament's ends are pinned to the $x$ axis and torque free: 
$u$ and $\p_x^2 u$ vanish at the endpoints. This choice enables one to expand the transverse 
undulations in a sine expansion
\beq 
\label{sine-expansion}
u(x,t) = \sum_p u_p(t) \sin(p x),
\eeq
with wave numbers 
\beq
\label{p-def}
p_n = n\pi/\ell,
\eeq 
where $n=1,2,\ldots$.

This Hamiltonian provides minimal coupling of a filament in a network to its surroundings (treated as a linear elastic solid). It is 
necessarily nonlinear.  The assumption of instantaneous tension propagation will eventually be violated at sufficiently 
high wavenumber since these undulatory modes will relax faster than the tension propagation time. 
Accounting for tension propagation introduces other nonlinearities to the Hamiltonian, which have 
been extensively studied~\cite{hallatschek2005,hallatschek2007a,hallatschek2007b}. We return to the 
relation of our work to these studies in Sec.~\ref{sec: MFT}.

The network is overdamped, being immersed in a viscous fluid with viscosity $\eta$ so that inertial effects may be ignored.  We treat the
hydrodynamic forces on the filament using resistive-force theory, where the drag force is linear in velocity and decomposes
locally into a component perpendicular to (with coefficient $\xi_\perp$) and parallel to (with coefficient $\xi_\parallel$) 
the mean tangent $\hat{t} \approx \hat{x}$. In terms of the position vector of a 
segment of the filament: $\vec u = (x, u_1(x),u_{2}(x))$, where the $1,2$ subscripts label the 
coordinates transverse to the direction of the undeformed filament $\hat{x}$, the drag force is~\cite{Wiggins1998}
\beq
\label{drag-force}
\left[\xi_\parallel \hat t \hat t + \xi_\perp (\mathbb{1} - \hat t \hat t)\right] \cdot \vec u= -\vec F_{\text{drag}},
\eeq
where the drag coefficients are given by $\xi_\perp \approx \frac{4\pi \eta}{\ln \ell/a}$, $\xi_\parallel \approx \xi_\perp/2$.  We neglect any modification of
the effective drag per unit length near the filaments ends, and we neglect any nonlocal hydrodynamics which 
produce logarithmic time corrections~\cite{Farge1993,granek1997R}.  The drag terms retained give the leading 
contribution to the drag forces in slender body theory, which provides a power series in $\ln(\ell/a)^{-1}$~\cite{Lighthill1976} at zero Reynolds number.  
Lastly, if we keep the drag forces acting on the filament only to linear order in $u$, we may neglect the drag associated with tangential motion.  

We now obtain overdamped, model A dynamics~\cite{Hohenberg1977}  
\beq
\label{model-A-dynamics}
\xi_\perp \p_t u(x,t) = - \delta H/\delta u(x,t) + \zeta(x,t).
\eeq
We also include Gaussian white noise 
\beq
\label{eq: noise correlations}
\langle \zeta(x,t) \zeta(x',t') \rangle = 2 \xi_\perp k_\text{B}T \delta(x-x') \delta(t-t')
\eeq
in the stochastic equation of motion, Eq.~\ref{model-A-dynamics}, consistent with the fluctuation-dissipation theorem. The analysis presented here is 
immediately generalizable to nonequilibrium and frequency-dependent noise, as long as it remains Gaussian.

\subsection{Spring-free results}
\label{sec: spring free}
We first review the previously studied dynamics of a filament with fixed applied tension and no coupling to springs. 
The Langevin equation is linear and admits a normal mode decomposition in terms of half integer wavelength sine waves discussed above.
Integrating over frequencies and averaging with respect to the white 
noise produces the dynamic correlation function for the amplitudes of these sine waves~\cite{Broedersz2014} -- see Eqs.~\ref{sine-expansion}, \ref{p-def}.
\beq
\langle u_p(t) u_p(0) \rangle = \frac{2k_\text{B} T}{\ell }\frac{e^{- \gamma_p^0 t/\xi_\perp}}{\gamma_p^0},
\eeq
where we have introduced
\beq
\label{eq: decay-rate}
\gamma_p^0 = \kappa p^{4} + \tau p^{2},
\eeq
so that $\gamma_p^0/\xi_\perp$ is the wavenumber-dependent decay rate. 
There are no cross correlations between amplitudes of different normal modes. 

There is a crossover between tension- and bending-dominated relaxational dynamics, 
set by the {\it tension length}
\beq
\label{eq: lt}
\ell_\tau = \sqrt{\kappa/\tau}.
\eeq
In the long-wavelength $\lambda \gg \ell_t$ tension-dominated regime, modes have an approximate relaxation time $\tau_{\text{relax}} \sim\frac{ \eta \lambda^2}{ \tau \ln(\ell/a)}$. In
the short-wavelength bending-dominated regime, modes have an approximate 
relaxation time $\tau_{\text{relax}} \sim \frac{ \eta \lambda^4}{\kappa \ln(\ell/a)}$. 
With vanishing applied tension, one observes a very broad range of relaxation times due to the  $\lambda^4$-dependence. We now consider dynamics with the
inclusion of the longitudinal spring, which mixes the filament's normal modes.

\section{The longitudinal spring: perturbative expansion}
\label{sec: perturbation theory}

We hereafter work in units such that $k_\text{B} T =1$. At the end of any calculation, we must then input factors of $k_\text{B}T$ where units of energy are needed. In these units, we can use the Einstein relation 
\beq
\label{eq: D-def}
D = \xi_{\perp}^{-1}
\eeq
to freely switch from $\xi_\perp$ to $D$, the latter of which represents a diffusion constant times a length. We now return to the full model A equation of motion defined by Eq.~\ref{model-A-dynamics}. By using Eqs.~\ref{eq:length} and~\ref{eq: H}, we find
\beq
\label{eq:nonlinear langevin}
\frac{\p u_p}{\p t} = - D \gamma_p^0 u_p - D k \Delta \ell u_p + h_p + \zeta_p,
\eeq
where $\zeta_p(t)$ and $h_p(t)$ represent noise and externally applied transverse force respectively, each absorbing a factor of
$D$.  From Eq.~\ref{eq: noise correlations}, we infer that equilibrium correlations of the Gaussian white noise 
obey the usual relation 
\beq
\langle \zeta_p(t) \zeta_{p'}(t') \rangle = (4 D/\ell) \delta_{pp'}\delta(t-t').
\eeq 

The second term on the right hand side of Eq.~\ref{eq:nonlinear langevin}, proportional to $k$, 
couples each mode (labeled by $p$) to changes in the total projected length of the filament, which depends on a sum over the square of amplitudes of 
all the dynamical modes.  As 
a result, this term in the equation of motion is nonlinear. In order to systematically compute correlation functions in the 
presence of this nonlinearity, we make use of the MSRJD functional integral method~\cite{tauber2014critical,altland2010}.

We start by introducing the moment generating MSRJD functional
\beq
\label{eq:Z-definition}
Z[j,\bar j] = \int \mathcal{D} [i\bar u(x)] \mathcal{D} [u(x)] e^{-\int \left(\mathcal{A}(\bar u, u)- \bar j \bar u - j u\, \right)dx dt},
\eeq
with the action $\mathcal{A}$ separated into: a Gaussian part $\mathcal{A}_0$, which generates correlation functions of the spring-free
system, the nonlinear and spring-dependent correction $\mathcal{A}_\text{int}$, and a term representing the external $h$-dependent forcing:
\beq
\label{Action}
\mathcal{A}[u(x,t),\bar u (x,t) ] = \mathcal{A}_0 + \mathcal{A}_\text{int} + D \int dx dt \, \bar u h.
\eeq
The Gaussian part is  
\beq
\label{Action-Gaussian}
\mathcal{A}_0 = \int dt dx  \left[\bar u \left( \p_t +  D (\kappa \p_x^4 - \tau \p_x^2) \right) u - D \bar u^2 \right],
\eeq
and the nonlinear interaction is
\beq
\label{Action-Nonlinearity}
\mathcal{A}_\text{int} = -\frac{D k}{2}\int dt dxdy \, \bar u(x,t) \frac{\p^2 u(x,t)}{\p x^2} \left(\frac{\p u(y,t)}{\p y}\right)^2.
\eeq
For the nonlinear action, we have explicitly written out the spatial and time dependencies. Each field is evaluated at the same time (a consequence of instantaneous tension propagation), yet there are two independent spatial variables $x$ and $y$ (nonlocality).

Finally, we recall that $(n,\bar n)$-point cumulants, representing response functions and correlation functions, are computed via functional derivatives
of the logarithm of the MSRJD functional:
\beq
\label{eq: functional derivative definition}
\langle \prod_{i,k}^{n,\bar n} u_i \bar u_k \rangle =\prod_{i,j}^{n,\bar n} \frac{\delta}{\delta j_i}\frac{\delta}{\delta \bar j_k} \ln Z[j,\bar j]|_{j=\bar j=0},
\eeq
where the brackets denote averages over the stochastic forces $\zeta(x,t)$. 
Specifically, by taking a derivative $\delta \langle u(x,t) \rangle/\delta h(x',t') |_{h(x',t')=0}$, we obtain the transverse linear response function:
\beq
\label{linear-response}
\chi_{uu}(x,x';t,t')=D \langle u(x,t) \bar u(x',t') \rangle.
\eeq
The source field $\bar j$ provides the same information as $h$. Hereafter we set $h=0$. The response function is trivially related to the propagator $G(x,x';t,t')$ of the theory via a factor of $D$:
\beq
\label{eq: G def}
G(x,x';t,t')=D^{-1}\chi_{uu}(x,x';t,t').
\eeq
We also define the dynamic or time-dependent correlation function
\beq
\label{eq: C def}
C(x,x';t,t') = \langle u(x,t) u(x',t') \rangle,
\eeq
hereafter referred to as the {\it correlator}. Given knowledge of $\chi_{uu}(x,x';t,t')$, it can be found easily via fluctuation-dissipation theorem, 
so it need not be calculated independently, at least for the equilibrium dynamics that we study here. 

We have chosen the Ito formulation of the Langevin equation, such that the Jacobian of our field transformation from $\zeta(x,t)$ to $u(x,t)$ is unity. 
This corresponds to the step function continuation $\Theta(0) =0$, and, as a result, all perturbative terms consisting of closed response loops 
evaluate to zero, consistent with causality. For general time ordering schemes, closed response loops can be shown to be canceled by the 
appropriate Jacobian factor, ensuring that the physical result is independent of discretization choice~\cite{tauber2014critical}.

\begin{figure}
\includegraphics[scale=1]{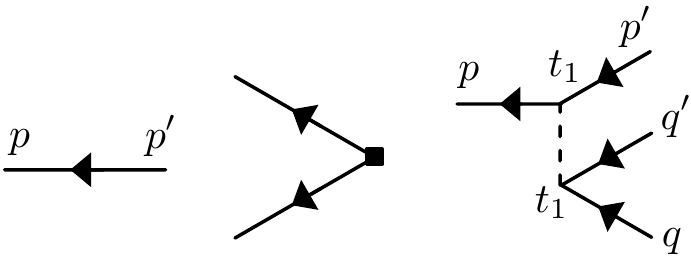}
\caption{Diagrams contributing to the perturbation theory of the $(u, \bar u)$ fields. The propagator (left) is a 
function of a single $p$ and the time difference: $\langle u_p(t) u_{p'}(t') \rangle = \delta_{pp'} G_p^0(t-t')$. The 
noise vertex (middle) produces two outgoing lines, and has a coefficient $D$. 
The interaction vertex (right) is equivalent to $\frac{-Dk\ell^2}{8} p^2 q^2 \delta_{pp'}\delta_{qq'}$. It carries two 
Kronecker deltas, and depends on two wavenumbers $p,q$. This is a consequence of the spatial nonlocality of the nonlinear interaction. 
Dashed lines connect two points at equal times. In the interaction vertex (right), 
we associated a factor of $p^{2} \delta_{p p'}$ ($q^{2}\delta_{q q'}$) with 
each vertex of the dashed and solid lines, and a factor of $\frac{-Dk\ell^2}{8}$ with the dashed line itself.}
\label{fig: vertices u}
\end{figure}

Eqs.~\ref{eq:Z-definition}, \ref{Action} enable the full machinery of diagrammatic perturbation theory in $k$. Expectation values with respect to the Gaussian 
action are denoted by the brackets $\langle \dots \rangle_0$. The diagrammatic rules are summarized in 
Fig.~\ref{fig: vertices u}. In wavenumber space, the Gaussian propagator is proportional to a Kronecker delta $\delta_{pp'}$, and therefore depends only on a single wavenumber. The retarded $(+)$ and advanced $(-)$ propagators are given by
\beq
\label{eq: G0}
G_p^{0,\pm}(t) = \frac{2}{\ell}\theta(\pm t) e^{ \mp \gamma_p^0 t},
\eeq
and represented by a directed line from earlier to later times. The comma in the superscript emphasizes that the $0$ is a label, and not related to the whether the propagator is advanced or retarded. The step function allows us to identify outgoing lines as $\bar u$ fields and incoming lines as $u$ fields. 

One may further define an undirected line to be the bare correlator
\beq
\label{eq: C0}
C^0(x,y,t) = \langle u(x,t) u(y,0) \rangle_0.
\eeq
However, since the bare correlator is related to the transverse linear response function via 
fluctuation-dissipation theorem, we can avoiding introducing the additional undirected propagator by treating the noise 
term $\sim D \bar u^2$ term in Eq.~\ref{Action-Gaussian} as a new vertex~\cite{tauber2014critical} 
denoted by the filled square in the middle of Fig.~\ref{fig: vertices u}. 

The spring-induced nonlinearity generates a spatially nonlocal, but instantaneous vertex.  As a result, the bare vertex shown in
Fig.~\ref{fig: vertices u} carries two independent Kronecker deltas in wavenumber and a delta function in time (dashed line), 
as well as four factors of wavenumber. See the caption of Fig.~\ref{fig: vertices u} for further details. We can easily switch from the time domain to the frequency domain, by Fourier transforming the fields
\beq
\label{eq: fourier transform}
u_p(t) = \int_{-\infty}^\infty \frac{d\omega}{2\pi} u_p(\omega) e^{-i \omega t},
\eeq
and imposing frequency conservation at each vertex.

Using these diagrammatic rules, we compute the $k$-dependent corrections to the propagator to two-loop order, which is also second order in $k$. 
Generally, in perturbation theory these corrections can be neatly grouped into a self-energy $\Sigma_p(\omega)$, defined by the relation
$\langle G_p(\omega) \rangle^{-1} = (G_p^0)^{-1}(\omega) - \Sigma_p(\omega)$~\cite{altland2010}. The physical interpretation of this quantity is found in the shift of the bare 
decay rate from Eq.~\ref{eq: decay-rate}, so that $\gamma_p^0 \to \gamma_p^0 -  \frac{2}{D \ell} \Sigma_p(\omega)$. As such, we define the adjusted self-energy
\beq
\label{eq: sigma tilde}
\tilde \Sigma_p(\omega) = \frac{2}{D \ell} \Sigma_p(\omega),
\eeq
which is precisely the shift in $\gamma_p^0$.

\begin{figure}
\includegraphics[scale=0.8]{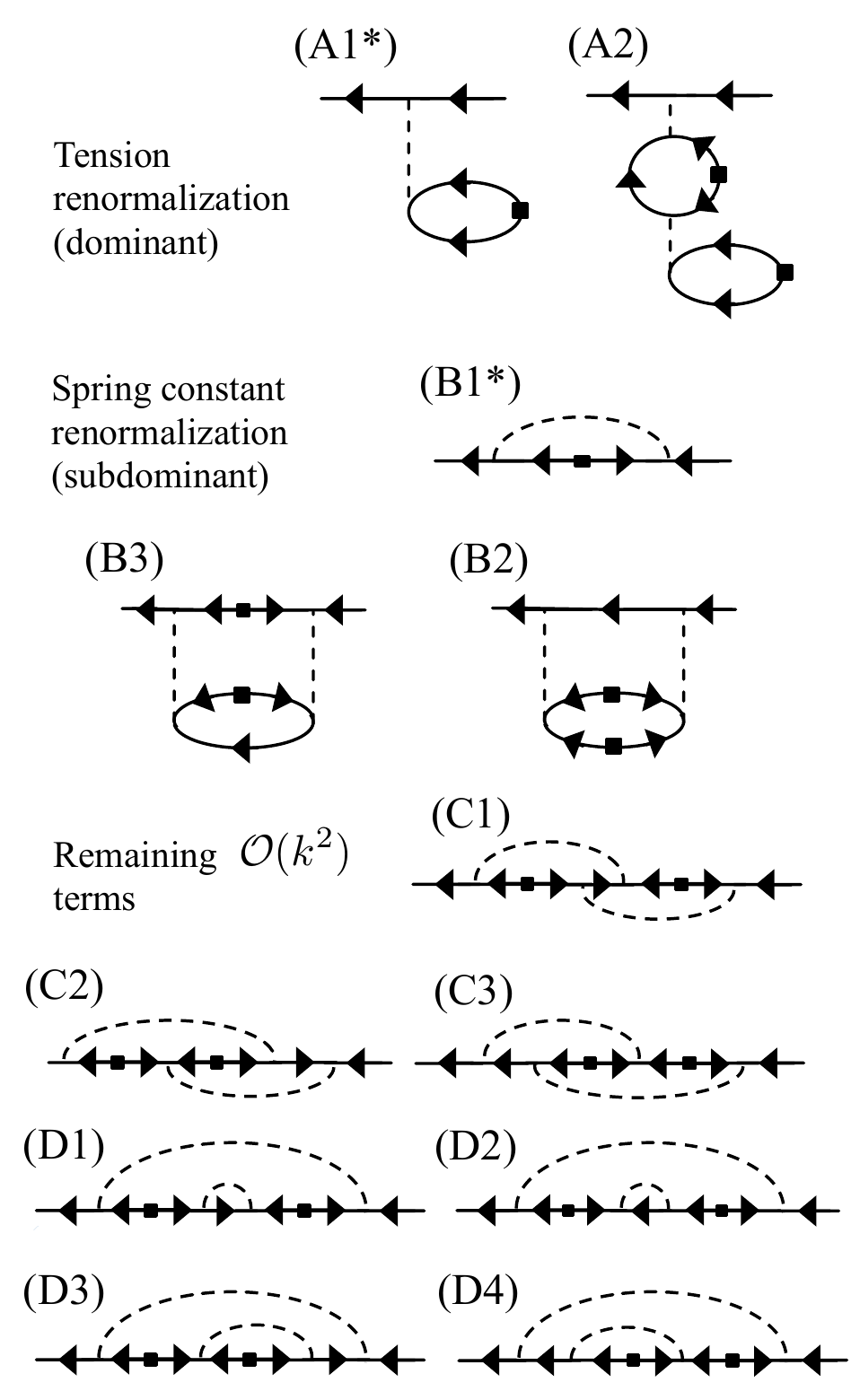}
\caption{All diagrams contributing to the self-energy (see definition preceding Eq.~\ref{eq: sigma tilde}) through $\mathcal{O}(k^2)$. 
There are two $\mathcal{O}(k)$ contributions marked by an asterisk. For detailed calculations, see Appendix~\ref{app: sigma}. 
Diagrams A1 and B1 are later used to renormalize tension (Eq.~\ref{eq: delta phi}) and self-consistently compute the 
self-energy (Eq.~\ref{eq: sigma nca def}).}
\label{fig: Ok2} 
\end{figure}

All the necessary diagrams for this calculation are shown in Fig.~\ref{fig: Ok2}, and we refer to them hereafter 
by their label in that figure, beginning with A1 at the top and continuing to D4 in the bottom right. They are individually calculated in Appendix~\ref{app: sigma}.
Here, we report the full two-loop self-energy (writing out $k_\text{B} T$ explicitly for clarity):
\begin{widetext}
\begin{eqnarray}
\label{eq: sigma o k2}
\tilde \Sigma_{\bar p}(\Omega) &=& -\frac{k k_\text{B}T \bar p^2}{\kappa} \left[ \frac{1}{\bar p^2 +1} + \frac{1}{2}\sum_{\bar q} \frac{1}{\bar q^2+1}\right] 
+ \frac{k^2 k_\text{B}^2T^2}{\kappa \tau^2}\bigg[ \frac{\bar p^2}{2(\bar p^2+1)^3}
+ \frac{3 \bar p^4}{(\bar p^2 +1)^2(-i \Omega + 3 \bar p^2(\bar p^2+1))} \nonumber \\
&+&\frac{1}{2} \frac{\bar p^2}{(\bar p^2+1)}\sum_{\bar q} \frac{1}{(\bar q^2+1)^2}\left(1 - \frac{-i\Omega}{-i \Omega + 2 \bar q^2(\bar q^2+1) + \gamma_{\bar p}} \right) 
+ \frac{\bar p^2}{4} \sum_{\bar q} \frac{1}{\bar q^2 +1}\sum_{\bar q} \frac{1}{(\bar q^2 +1)^2} \bigg].
\end{eqnarray}
\end{widetext}
We have introduced dimensionless wavenumbers
$ \bar p = p \sqrt{\kappa/\tau} = p \ell_\tau$  and frequencies $\Omega = \frac{\omega \kappa}{D\tau^2}=\omega/\omega^*$. 
These units are convenient, provided that the tension is not so small that $\ell_\tau \approx \ell$, but they are primarily used in order to aid in a qualitative analysis of Eq.~\ref{eq: sigma o k2}. For an alternative scheme valid at small $\tau$, see Eq.~\ref{eq: sigma phi}.

The $\mathcal{O}(k)$ correction is negative, and proportional to $\bar p^2$, which tells us that it renormalizes the effective tension to 
a larger value. This is expected, as the spring stiffens the filament to elongation, causing it relax faster. 
To analyze the effects of the spring beyond first order, we categorize the two-loop self-energy into 
three types of contributions, beginning with the most dominant. The first group consists of type A diagrams in Fig.~\ref{fig: Ok2}.
These are $\sim \bar p^2$ so they become appreciable at large wavenumber. Due to the summations, they also grow with system size, {\em i.e.}, filament length. 
As we will see in Sec.~\ref{sec: MFT} (and commented on more in Appendix~\ref{app: sigma}), these correspond to a renormalization of the tension, and 
may be eliminated by using a self-consistent approach to the Green's function.

The second group consists of the type B diagrams in Fig.~\ref{fig: Ok2}. After $\tau$ renormalization, these are the next most important class of diagrams. We will later find that they correspond to renormalization of the spring constant $k$. At large $\bar p$, they plateau to a constant value, and, at small $\bar p$,
they decay as $\bar p^2$. These corrections are important for $\bar p \le 1$. These contributions are largest at zero frequency, where they acquire 
a prefactor $\sim \sum_{\bar q} (\bar q^2+1)^{-2}$. But this remains small when compared to the type A diagrams, which are proportional 
to $\sum_{\bar q} (\bar q^2 +1)^{-1}$. In general, we will find (see Sec.~\ref{sec: fluctuations}) that any diagram containing a 
solid loop with $n$ outgoing dashed lines will be proportional to a summation $\sum_{\bar q} (\bar q^2 +1)^{-n}$, and thus 
represent increasingly smaller contributions.

The third and final group consists of both type C and D diagrams of Fig.~\ref{fig: Ok2}. These diagrams have a 
single solid line with crossed (type C) or uncrossed (type D) dashed lines. At large $\bar p$, these vanish and are 
therefore small compared to the diagrams of the first (A) and second groups (B). At small $\bar p$, they go to zero as $\bar p^2$, however, they lack a 
summation compared to the other terms in Fig.~\ref{fig: Ok2} and are thus still smaller. At $\mathcal{O}(k^2)$, 
these summations are $\sim \sum_{\bar p} \bar p^{-2}$. As a result, we infer that the missing summations in type C and D diagrams cause them to be about 
an order of magnitude smaller than the contributions from the other $\mathcal{O}(k^2)$ diagrams. Furthermore, at high frequency, the 
contributions from the crossed (C) diagrams are smaller than those from non-crossing (D) diagrams. This suggests that we
may ignore crossed diagrams in any self-consistent treatment of the dynamics, as described below.  This distinction between the crossing and non-crossing
diagrams is analogous to impurity scattering in condensed matter, where one also finds that crossing diagrams in electron impurity scattering calculations 
may be safely ignored~\cite{rammer2018,ziman1979}. 

We now use the previous analysis to develop a self-consistent approximation for the propagator of Eq.~\ref{eq: G def} in frequency/wavenumber space. 
The principal effect of the longitudinal spring is to renormalize tension. The details of that process will be shown in Sec.~\ref{sec: MFT}. 
We account for this by defining
\beq
\label{eq: gamma def}
\gamma_p = \kappa p^4 + \tau_\text{R} p^2,
\eeq
which everywhere replaces $\gamma_p^0$. $\tau_\text{R}$ is the renormalized tension due to the longitudinal spring. We 
next incorporate the remaining first order correction (diagram A1), by considering it as the first term in a series of diagrams that contain a single solid line, 
with no crossed dashed lines (the $\mathcal{O}(k^2)$ term in this series consists of all type D diagrams in Fig.~\ref{fig: Ok2}). The infinite 
summation can quickly be achieved by demanding that the self energy is equal to the contribution in diagram A1, so long as we replace the 
bare propagators by a dressed ones. This leads to the self-consistent equation
\beq
\label{eq: sigma nca def}
\tilde \Sigma_p^\text{NCA}(\omega) = - \frac{k k_\text{B} Tp^4}{\gamma_{p}-\tilde \Sigma_p^\text{NCA}(\omega)},
\eeq
known as the non-crossing approximation (NCA). This is certainly correct to $\mathcal{O}(k)$, and as $\omega \to \infty$ 
becomes precise to all orders in $k$. Since this is a self-consistent equation, we are free to extend $k$ to large values where we can 
see its effect. Eq.~\ref{eq: sigma nca def} is algebraic, and we easily find the solution
\beq
\tilde \Sigma_p^\text{NCA}(\omega) = \frac{\gamma_p}{2} \left(1 - \sqrt{1+\frac{4 k k_\text{B} T  p^4}{\gamma_p^2}}\right).
\eeq
The simplicity of this result is a direct consequence of the spatial nonlocality of our interaction; since dashed lines do not carry wavenumber, there is no summation over modes in diagram A1. From $\tilde \Sigma^\text{NCA}_p(\omega)$, we find the NCA transverse linear response function
\beq
\label{eq: chi nca}
\chi^\text{NCA}_{p}(\omega) = \frac{2D/\ell}{- i \omega + \frac{1}{2}D\gamma_p \left(1 + \sqrt{1 + 4k k_\text{B} T p^4/\gamma_p^2}\right)}
\eeq
Using the fluctuation-dissipation theorem and reinserting $k_\text{B} T$ where necessary to work in physical units, we
obtain the dynamic correlator
\beq
\label{eq: C nca}
C^\text{NCA}_p(\omega) = \frac{4 k_\text{B}T/\xi_\perp \ell}{\omega^2 + \frac{\gamma_p^2}{4 \xi_\perp^2} \left(1+\sqrt{1+ 4 k k_\text{B}T p^4/\gamma_p^2}\right)^2}.
\eeq

\begin{figure}
\includegraphics[scale=0.63]{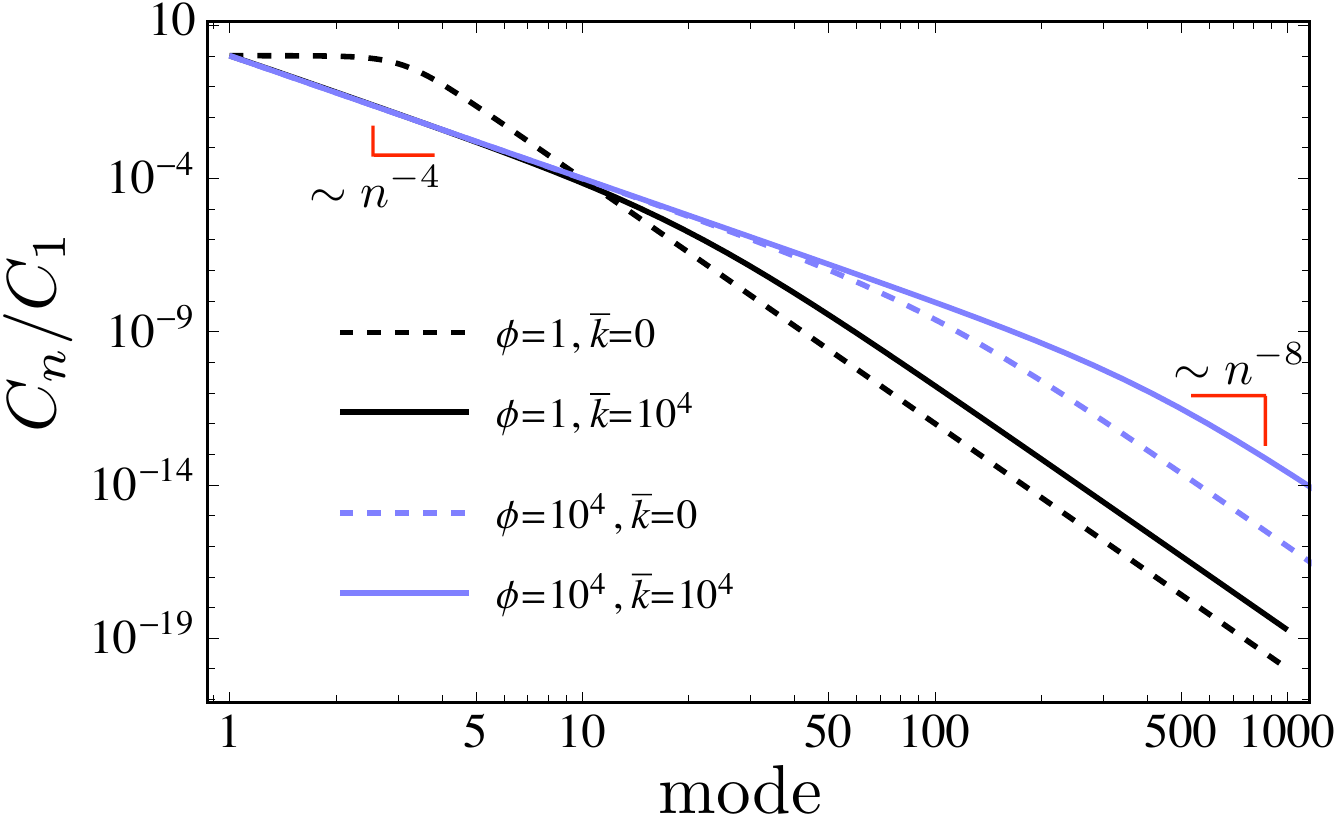}
\caption{(color online) NCA dynamical correlation function normalized by its first mode at both low (black) and high (blue) tension in the presence (solid)
or absence (dashed) of the longitudinal spring. $\bar \omega =100$. The solid black curve overlaps with the solid blue curve at low mode numbers, 
indicating that the spring generates tension in the absence of any pre-existing tension, given in a nondimensionalized form as $\phi$. In the presence 
applied tension $\phi>0$, the spring increases the effective tension, pushing the transition from tension- to bending-governed fluctuations to higher mode 
numbers (blue curves).}
\label{fig: CvsMode}
\end{figure}
At low tension, we intuitively expect the effect of the longitudinal spring to be stronger. We thus seek units in which the tension 
can easily be taken to small values. Per the discussion of Sec.~\ref{sec: spring free}, at low wavenumber, the system 
is in a tension-dominated regime. Accordingly, we switch to a dimensionless length scale by factoring out the wavenumber of 
the lowest mode $p_1=\pi/\ell$.   We also adopt a dimensionless tension, spring constant, and frequency:
\begin{subequations}
\label{eq: dimensionless quantities}
\beq
\label{eq: phi def}
\phi = \frac{\tau \ell^2}{\kappa \pi^2},
\eeq
\beq
\label{eq: k bar def}
\bar k = \frac{k k_\text{B} T \ell^4}{\kappa^2 \pi^4},
\eeq
\beq
\label{eq: omega bar def}
\bar \omega = \frac{ \omega \xi_\perp \ell^4}{\kappa \pi^4}.
\eeq
\end{subequations}
To compute $C_p(\omega)$, we must further calculate tension renormalization. In terms of $\phi$, this amounts to the replacement
\beq
\phi_R = \phi + \Delta \phi,
\eeq
where $\Delta \phi$ is defined by the self-consistent equation
\beq
\label{eq: delta phi}
\Delta \phi = \frac{ \bar k }{2} \sum_{n=1}^\infty \frac{1}{n^2 + \phi + \Delta \phi}.
\eeq
This equation can be derived by approximating the entire self-energy correction by the dominate diagram A2 in Fig.~\ref{fig: Ok2}, 
provided we replace the loop propagator with the dressed one. This approximation is discussed more fully in the context of the mean-field theory 
in Sec.~\ref{sec: MFT}. In terms of mode number $n$, we find the correlator to be
\beq
C_n^\text{NCA}(\bar \omega) = \frac{ 4k_\text{B} T\xi_\perp \ell^7/\kappa^2 \pi^8}{\bar \omega^2 + \left(\frac{1}{2} n^2(n^2+\phi_R)\left(1+\sqrt{1+\frac{4 \bar k}{(n^2+\phi_R)^2}}\right)\right)^2}.
\eeq

In Fig.~\ref{fig: CvsMode}, we plot the NCA correlator as a function of wavenumber.   Generically there are three regimes going from low to high 
mode number. There is a low wavenumber plateau transitioning into a $n^{-4}$ decay, followed by an $n^{-8}$ decay at sufficiently high mode numbers.  The 
effect of the spring is to shift these transitions to lower mode number.  For sufficiently high spring constants, the plateau regime may disappear 
entirely as shown by the (blue and black) solid curves in the figure.  The condition for the appearance of the plateau is that $\bar \omega > \max\{n^4/4,\phi/4\}$
for some $n \ge 1$.  The principal effect of the spring is still tension renormalization. Even as 
$\phi \to 0$, the longitudinal spring ensures that the filament still behaves as if it were under tension.  For finite values of the applied tension, 
the effect of the spring still increases the total effective or renormalized tension, moving the transition to higher-frequency,  bending-dominated 
fluctuations to still higher modes.  The fact that the mode where the fluctuations change from being stretching- to bending-dominated
moves in response to the external spring suggests that the 
effects of even a weak spring will be most easily observed near this tension-to-bending transition ($p= \ell_\tau^{-1}$) of the spring-free model.

\begin{figure}
\includegraphics[scale=0.6]{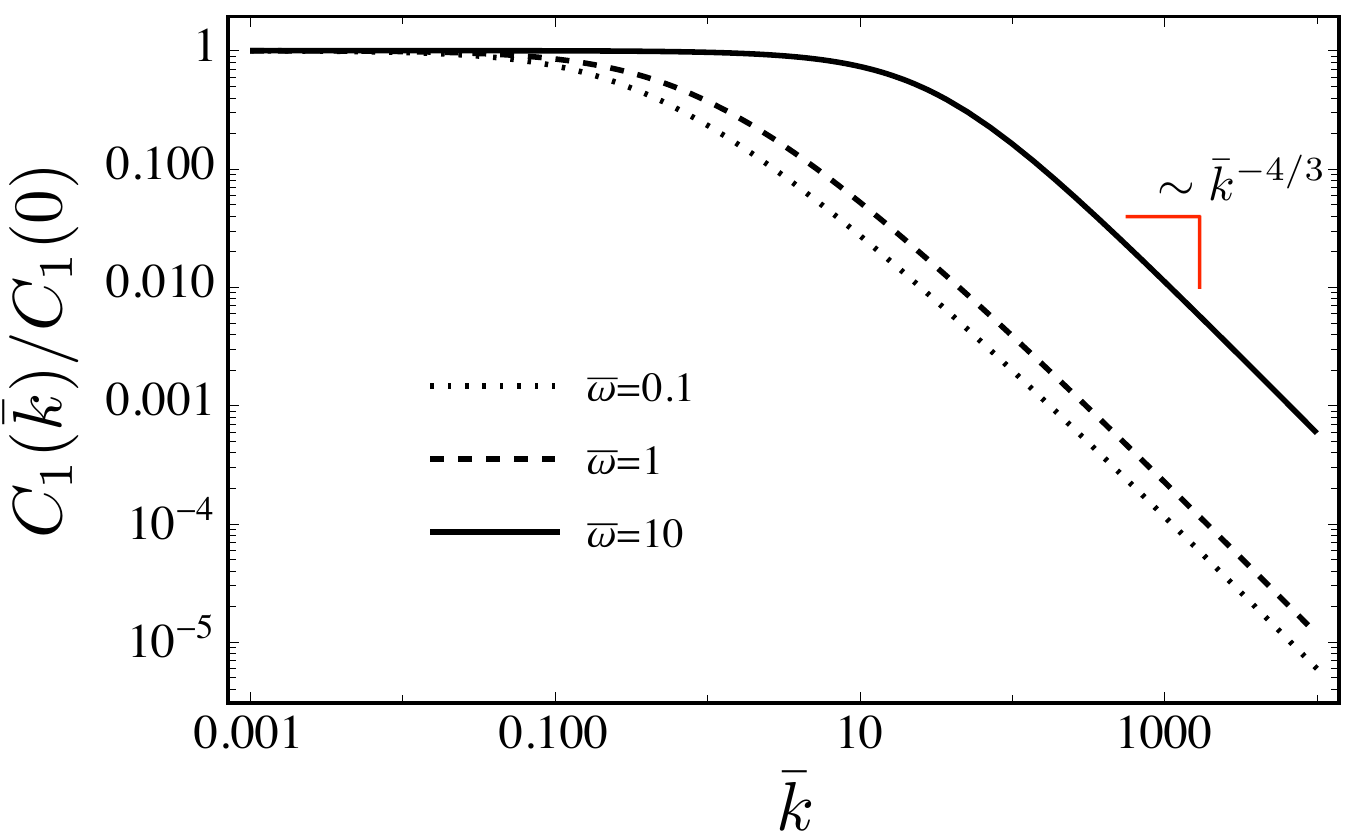}
\includegraphics[scale=0.6]{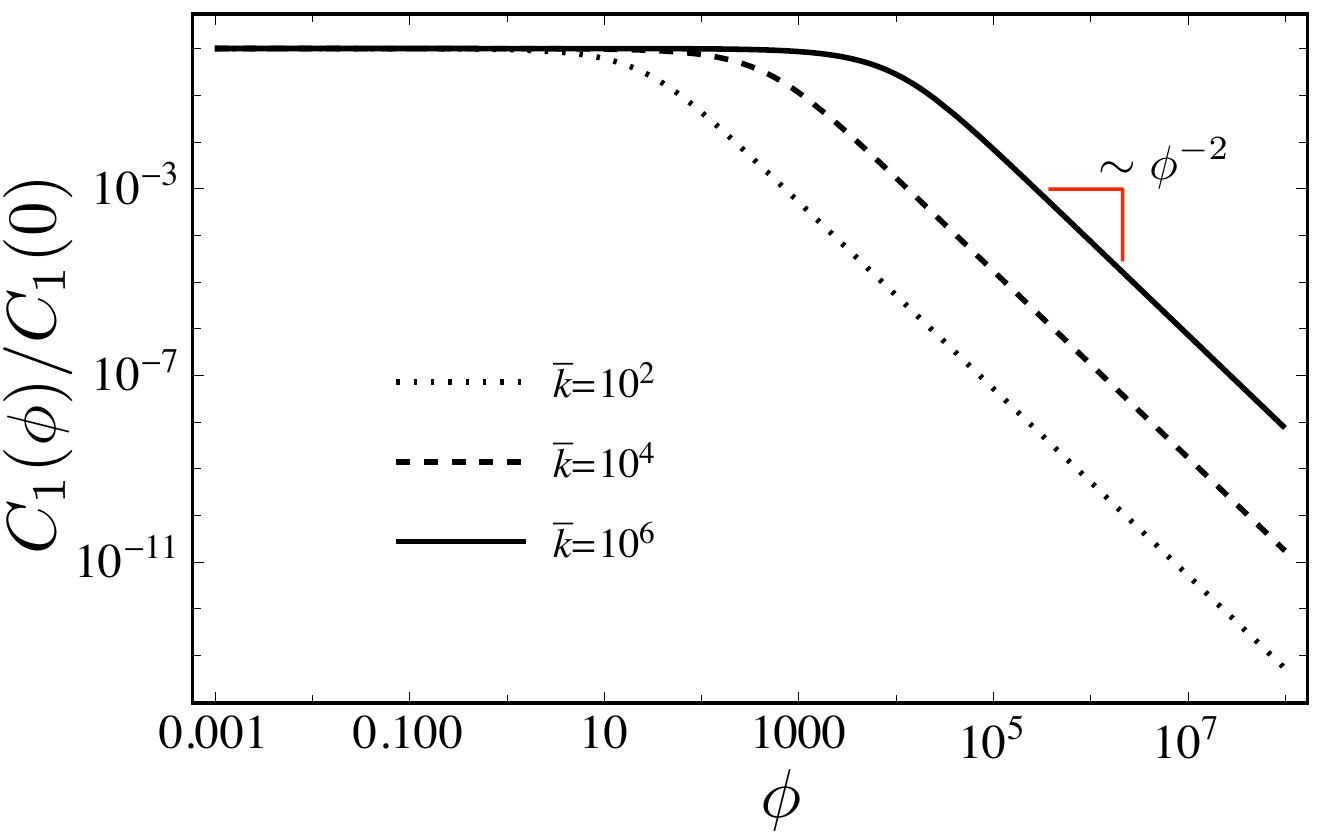}
\caption{(color online) Lowest mode of the dynamic correlation function vs.~(top) spring constant and (bottom) applied tension. The top panel is 
evaluated at low tension, $\phi=10^{-2}$, and the bottom at $\bar \omega =1$. At large $\bar k$, the effective tension grows sublinearly 
as $\sim \bar k^{2/3}$, leading to the $\bar k^{-4/3}$ dependence of $C_1$. In the bottom panel, the lowest mode dynamic 
correlation function decays as $\phi^{-2}$, which is identical to the spring-free $k=0$ case. The transition to the $\phi^{-2}$ decay occurs at 
tensions higher than $\phi \approx \bar k \langle \Delta \ell \rangle$.}
\label{fig: CvsPhiK}
\end{figure}
In Fig.~\ref{fig: CvsPhiK}, we look at how varying the applied tension and spring constant shift the lowest mode $C_1(\omega)$. If the spring does 
not significantly alter $\Delta \ell$, then the tension $k \Delta \ell$ created by the spring increases linearly in $k$.
However, due to the self-consistent condition, at high $k$, $\Delta \ell$ diminishes, causing tension to increase as $k^{2/3}$. We 
discuss this scaling more fully in Sec.~\ref{sec: MFT}.  As a result, the correlator decays like 
$\bar k^{-4/3}$, as shown in the top panel of Fig.~\ref{fig: CvsPhiK}. The transition occurs when $k^*= \tau/\Delta \ell$, which we 
approximate as $k^* \approx 12 \kappa \tau/k_\text{B}T \ell^2$ by replacing $\Delta \ell$ with its small tension and spring-free result~\cite{mackintosh1995}.

We can also see the transition in the correlation function by keeping $k$ constant and varying tension.  The correlator transitions from being 
$\phi$-independent to decaying as $\phi^{-2}$ with increasing $\phi$, as shown in the bottom panel of Fig.~\ref{fig: CvsPhiK}. The dependence of 
the correlator upon applied tension is the 
same as in the spring-free model. The transition occurs once $\phi$ is greater than both $ 4\bar\omega^2$ (for the lowest mode)
and the renormalized tension $\Delta \phi$, due to the spring.   As a result, the spring washes out the effect 
of small applied tensions, replacing the overall tension with its renormalized value. We now turn to a justification of the 
approximations outlined above, as well as derive new results concerning projected length fluctuations. 
Our main tool will be functional techniques using the MSRJD formalism.

\section{Projected length auxiliary field theory}
\label{sec: auxiliary field theory}
The spatially nonlocal theory presented here was previously examined in equilibrium, where the nonlocal aspect allowed for a 
complete resummation of diagrams contributing to the two point function~\cite{kernes2020equilibrium}. In the dynamical version, however, this 
resummation fails. The previous calculation of equal-time correlation functions allowed for a great simplification due to the fact 
that all of these diagrams collapsed into one of two groups - see Ref.~\cite{kernes2020equilibrium}. The calculation of 
dynamical correlations here, however, introduces a time associated with each interaction. This time 
ordering makes all the previously identical diagrams from Ref.~\cite{kernes2020equilibrium} distinct. 
Since, in the dynamical theory, dashed lines carry frequency, there are an infinite number of inequivalent single-line diagrams, 
differentiated by the arrangement of dashed-line contractions (for example, compare the class C and D diagrams in Fig.~\ref{fig: Ok2}). 

Despite this complication, we may still proceed along the lines of Ref.~\cite{kernes2020equilibrium}. Inspection of Eq.~\ref{eq: H} 
suggests that the Hamiltonian is more naturally expressed in terms of $\Delta \ell(t)$ rather than $u(x,t)$. This will allow us to more easily 
compute projected length fluctuations $\langle \Delta \ell(t) \Delta \ell(t') \rangle$, which are relevant for experiments measuring the dynamic shear modulus. 
As a tradeoff, solving for the two-point function, $\langle u_p(\omega)u_{p'}(\omega')\rangle$, will be harder.

\begin{figure}
\includegraphics{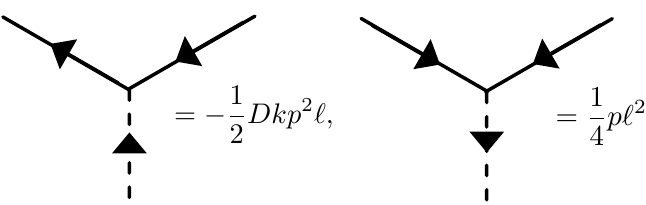}
\caption{$\lambda \bar u u$ and $\bar \lambda u u$ interactions. The Hubbard-Stratonovich transformation cuts the four-point vertex 
into two three-point vertices. Dashed lines are now directed, with $\lambda$ incoming and $\bar \lambda$ outgoing.}
\label{fig: lambda u u}
\end{figure}

In order to change functional integration variables from $u(x,t) \to \Delta \ell(t)$, we first employ a Hubbard-Stratonovich 
transformation to write the quartic interaction, $\frac{1}{2} k \Delta \ell^2$, in terms of an interaction with auxiliary fields $\lambda, \bar \lambda$. 
This amounts to using the identity~\cite{zinn1996quantum}
\beq
\label{eq: HS identity}
e^{-\int dt \bar z z } = \int \mathcal{D}(\bar \lambda,\lambda) e^{-\int \left[ \bar \lambda \lambda - \bar z \lambda - \bar \lambda z  \right]dt }
\eeq
in Eq.~\ref{eq:Z-definition}, while making the identifications $\bar z = -D k \int \bar u u'' dx $ and $z= \Delta \ell=\frac{1}{2} \int u'^2 dx$. 
Diagrammatically, this transformation severs the undirected dashed line into the two three-point vertices depicted in Fig.~\ref{fig: lambda u u}.
This transformation is essentially a $\delta$-function, acting to assign the change in projected length to the variable $\lambda(t)$. We may alternatively arrive at this step by introducing a Lagrange multiplier into the Hamiltonian, writing down the Langevin equation, then finding the MSRJD functional.

We now add additional source terms $\int dt j_\lambda(t) \lambda(t) + \bar j_\lambda \bar \lambda(t)$ to the expanded functional, which 
will generate correlations of the auxiliary $\lambda, \bar \lambda$ fields. To understand the physical meaning of these new auxiliary fields, we take 
the functional derivatives $\frac{\delta}{\delta j_\lambda},\frac{\delta}{\delta \overline{j}_{\bar \lambda}}$ of the generating functional before and 
after integration over $(\bar \lambda, \lambda)$, and compare the results. $\delta Z[j_\lambda, \bar j_{\bar \lambda}]/\delta j_\lambda$ produces the 
moments of $\Delta \ell(t)$. As a result, there is a one-to-one correspondence between expectation values of $\lambda(t)$ and $\Delta \ell(t)$. 
That is, for any $N$-point correlation
\beq
\label{eq: lambda to deltaL}
\langle \prod_i^N \lambda(t_i) \rangle = \langle \prod_i^N \Delta \ell(t_i) \rangle.
\eeq

$\bar \lambda$ is related to the linear response of the projected length to an applied tension. 
For a small change in applied tension $\Delta \tau$, this is defined as
\beq
\chi_{\Delta \ell}(t,t') = \left. \frac{\delta \Delta \ell(t)}{\delta \Delta \tau(t')}\right|_{\Delta \tau=0}.
\eeq
Eq.~\ref{eq: HS identity} shows that $\bar \lambda$ appears conjugate to $\Delta \ell$, in the same manner as would a time-dependent applied tension. 
 Applying two derivatives $ \left. \frac{\delta^2\ln Z}{\delta j_\lambda \delta \bar j_{\bar \lambda}}\right|_{j_\lambda=\bar j_{\bar \lambda} =0}$ before and after integration over auxiliary fields, and then comparing the results, we find the linear response is expressed in terms of the auxiliary fields as

\beq
\label{eq: chi lambda}
\chi_{\Delta \ell}(t,t') = k^{-1}(1 - \langle \lambda(t) \bar \lambda(t')\rangle).
\eeq

As a result of the Hubbard-Stratonovich transformation, the action $\mathcal{A}$ now depends on four fields $\mathcal{A}[\bar \lambda,\lambda,\bar u,u]$. 
It is quadratic in the fields $\bar u, u$, so we may integrate them them out. Doing so, yields the effective action
\begin{subequations}
\beq
\label{eq: A lambda}
\mathcal{A}[\bar \lambda, \lambda,\bar j, j] = \int \bar \lambda \lambda dt + \frac{1}{2} \text{Tr} \ln {\bf G}^{-1} - \frac{1}{2} \int {\bf j}^T {\bf G} {\bf j} dt,
\eeq
\beq
{\bf G}_p^{-1} = \left(\begin{array}{cc} -2 D \mathbb{1} & (G_p^+)^{-1} \\ (G_p^-)^{-1} & - p^2 \bar \lambda \mathbb{1} \end{array}\right)
\eeq
\end{subequations}
plus $\lambda, \bar \lambda$-dependent source terms. In the above expressions, the lower case, bold letters stand for the vectors of the fields
${\bf u} =(\bar u,u)$, $\bm{\lambda}=(\bar \lambda, \lambda)$, and ${\bf j} = (\bar j, j)$. The trace runs over fields. It also includes a 
summation over wavenumbers $p$.  We have defined the 2x2 block matrix (since its components are operators) ${\bf G}^{-1}$ in terms of the 
advanced/retarded propagators
\beq
\label{eq: G plus definition}
G_p^\pm(t,t')=G_p^{0,\pm}(t) e^{ \mp D k p^2 \int_{t'}^t \lambda(t'') dt''}.
\eeq

Since $\mathcal{A}$ retains its dependence on the source terms $\mathbf{j}$, we may still generate correlations of the transverse 
displacement field via functional differentiation, as defined in Eq.~\ref{eq: functional derivative definition}. As expected, correlations 
$\langle \mathbf{u}\mathbf{u} \rangle$ depend on expectation values of operator inverses containing stochastic fields $\bm{\lambda}$. We 
have traded calculating a simple observable with a complex probability functional for a nonlinear observable 
with a simple probability functional. Correlations with respect to $\Delta \ell$, on the other hand, are evaluated at $\mathbf{j}=0$ and are tractable, 
provided we can simplify the trace-log appearing in $\mathcal{A}$.

Since the spring constant $k$ appears only in the combination $\sim D k p^2 \lambda$, we may shift integration variables 
$\lambda \to \lambda/D k$, thereby putting all of the $k$ dependence in $\mathcal{A}$ into the first term $\int dt \bar \lambda \lambda /D k$. 
As $k\to0$, $\mathcal{A}$ oscillates wildly, indicating that saddle-point evaluation of the functional integral becomes exact. We may then carry out a controlled small $k$ expansion of $\mathcal{A}$ about its saddle-point solution $(\bar \lambda_0, \lambda_0)$ plus fluctuations. Incidentally, 
the saddle-point solution $\lambda_0$ is {\it precisely} the average $\langle \Delta \ell(t) \rangle$, regardless of whether or not $k$ is small.

\subsection{Mean field theory}
\label{sec: MFT}

We investigate the saddle-point solution corresponding to the effective action Eq.~\ref{eq: A lambda}, which becomes exact as $k \to 0$. 
We denote the saddle point solutions for the auxiliary fields by $\lambda_0$ and $\bar \lambda_0$. We will find that the saddle-point solution 
corresponds to a type of dynamical ``mean-field theory'' (MFT),
and henceforth refer to $\bar \lambda_0, \, \lambda_0$ as the mean-field solutions. 

The saddle-point equations are
\beq
\delta \mathcal{A}/\delta \lambda= \delta \mathcal{A}/\delta \bar\lambda =0,
\eeq
evaluated at $\lambda=\lambda_0$ and $\bar \lambda= \bar \lambda_0$. Functional differentiation of the trace-log appearing 
in $\mathcal{A}$ is carried out in the standard way~\cite{altland2010}, using 
$\delta_{\bar \lambda} \Tr \ln \mathbf{G}^{-1} = \Tr \left( \mathbf{\hat G} \delta_{\bar \lambda} \mathbf{\hat G}^{-1} \right)$. 
As $\bar \lambda$ appears only in the (22) component of $\mathbf{G}^{-1}$, functional differentiation yields a matrix with one in the (22) 
component, and zeroes elsewhere. Taking the matrix product with $\mathbf{G}$ and performing the trace yields the $(22)$ component of 
$\mathbf{G}$. We emphasize again that $\mathbf{G}^{-1}$ is really a 2x2 block matrix, with each block representing an operator. 
Since  $\mathbf{G}^{-1}$ is not diagonal in either the time or frequency domains, we cannot trivially invert it. Instead, we determine 
$\mathbf{G}$ via its defining equation $(\mathbf{\hat G}^{-1})_{ik}  \mathbf{\hat G}_{kj} = \delta_{ij} \delta(t-t')$. 
This yields the result $G_{22} = (\mathbb{1} - p^2 \hat C_p \bar \lambda_0)^{-1} \hat C_p$. 

Since $\delta \mathbf{\hat G} /\delta \lambda=0$, the first saddle-point equation is trivially
\beq
\bar \lambda_0 (t)=0.
\eeq
The second saddle-point equation can now be easily found by setting $\bar \lambda_0=0$. We find the second saddle-point equation
\beq
\label{eq: MFT definition}
\lambda_0(t) = D \sum_p p^2 \int_{-\infty}^t \left[G_p^+(t,t')\right]^2 dt',
\eeq
where $G_p^+(t,t')$ was defined in Eq.~\ref{eq: G plus definition}. This depends only on $\lambda_0$, and we call it the mean-field condition.

There are two alternative ways to interpret this result, each of which add to our physical understanding. 
First, in the context of the $(\bar u, u)$ diagrammatic perturbation theory defined by Fig.~\ref{fig: vertices u}, we can recover the 
mean-field condition by summing over all one-correlator loop corrections to the propagator. These contributions can be grouped into a 
mean-field self energy $\Sigma^\text{MFT}_p(t)$. We then demand that $\Sigma^\text{MFT}_p(t)$ is equivalent to diagram A1 in Fig.~\ref{fig: Ok2}, 
when the loop correlator is replaced by a dressed correlator. Looking for a solution of the form 
$\tilde \Sigma^\text{MFT}_p(t) = -\bar k p^2 \lambda_0$ reproduces the mean-field condition.
This observation suggests that the mean-field theory is the leading term in an expansion of $\mathcal{A}[\bar \lambda, \lambda]$, 
determined by the maximal number of dashed lines emanating from a closed, solid line loop. We call a subdiagram with $n$ outgoing 
dashed lines an {\it n-bubble}. The suggestion turns out to be accurate, and is elaborated on more in Sec.~\ref{sec: fluctuations}. 

Second, we may arrive at Eq.~\ref{eq: MFT definition} by employing a type of {\it mean-field} approximation, in which we make the
replacement:  $\Delta \ell^2 \to 2 \langle \Delta \ell \rangle \Delta \ell$ in the Hamiltonian -- see Eq.~\ref{eq: H}.  The 
angled brackets denote averages with respect to the noise.  Looking at this replacement more closely, we note that the equilibrium average
$\langle \Delta \ell(t) \rangle$ must be a constant in time. 
Here, however, the averaging is applied with respect only to the noise, and not to the initial configuration of the filament. In that case, 
the average $\langle \Delta \ell(t) \rangle$ can evolve in time from any particular initial condition. The mean-field theory is 
capable of describing the relaxation of this variable to its equilibrium value. For example, we can consider a situation where the filament is 
pulled starting at time $t=0$.  

Returning to our mean field approximation, the MFT Hamiltonian is now linear. The resulting Langevin equation is also linear, 
and can be solved for $u_p(t)$ in terms of the noise $\zeta_p(t)$ and $\langle \Delta \ell(t) \rangle$. Imposing the self-consistency 
condition given by the definition in Eq.~\ref{eq:length} of projected length, and identifying $\lambda_0(t) = \langle \Delta \ell(t) \rangle$, we 
reproduce the mean-field condition Eq.~\ref{eq: MFT definition}. Physically, the mean-field approximation assumes that the normal modes 
respond only to the change in the averaged projected length, and ignore changes in $\Delta \ell$ due to fluctuations of other normal modes. 
This approach is actually a mean-field {\it differential} equation for the function $\lambda_0(t)$. The mean-field theory is neatly 
summarized as the following Langevin equation
\begin{subequations}
\label{eq: MFT equations}
\beq
\label{eq: MFT eq1}
\frac{\p u_p(t)}{\p t} = - D \left[ \gamma_p^0 +k p^2 \lambda_0(t) \right] u_p(t) + \zeta_p(t) ,
\eeq
with the condition
\beq
\label{eq: MFT eq2}
\lambda_0(t) = \frac{\ell}{4} \sum_p p^2 \langle u_p^2(t) \rangle.
\eeq
\end{subequations}

Combining these two equations results in the integral equation given by Eq.~\ref{eq: MFT definition} for $\lambda_0(t)$. When 
solving the integral equation, it is more convenient to work with the time derivative of $\lambda_0(t)$. Following the notation of 
Hallatschek {\em et al.}~\cite{hallatschek2005,hallatschek2007a,hallatschek2007b}, we hereafter refer to Eq.~\ref{eq: MFT definition} 
as a partial-integro differential equation (PIDE). The quantity $k \lambda_0(t)$ acts as a time-dependent tension, 
whose value depends self-consistently on the instantaneous conformation of the filament. Our model appears similar to those 
describing nonlinear tension propagation along inextensible filaments~\cite{seifert1996,hallatschek2005,hallatschek2007a,hallatschek2007b}. 
This is true for both the ordinary and
multiscale perturbation theory~\cite{hallatschek2007a}. These authors obtain a PIDE similar to ours, where our $\lambda_0(t)$ is 
analogous to their stored thermal length $\langle \varrho(t) \rangle$.
Our analysis differs from the previous work in that the inherent longitudinal compliance of the system is 
concentrated in the external longitudinal spring, rather than the extensional deformation of the filament. The 
longitudinal spring responds only to a particular, collective degree of freedom 
of the system -- the end-to-end length. Moreover, the longitudinal spring constant can be changed arbitrarily for a filament with fixed elastic compliance, which
provides more freedom for exploration.

When comparing our analysis to the multiscale perturbation theory PIDE, the key distinction is that 
our $\lambda_0(t)$ does not have spatial dependence. Theories of tension propagation in untensed and 
tensed filaments~\cite{seifert1996,hiraiwa2008,obermayer2009tension} allow for a finite propagation speed of tension, which 
requires that the longitudinal extension be spatially dependent.  
In either case, provided we are looking at filaments short enough that we may neglect the finite
speed of tension propagation, our results should hold.

In Eq.~\ref{eq: spring free chi}, we show the predicted response function of the end-to-end distance in the presence of prestress.  
While the projected length fluctuations in the absence of prestress have 
been studied~\cite{gittes1998G,morse1998visc,granek1997R}, there has not been an explicit discussion of the problem with 
prestress~\footnote{The response of unstressed filaments in an equilibrium ensemble to an applied tension at $t=0$ has been studied~\cite{granek1997R,hallatschek2007a},
and incorrectly extended to filament fluctuations under prestress via fluctuation dissipation theorem~\cite{Broedersz2014}. The response is not in the linear forcing regime, and thereby the fluctuation dissipation theorem does not hold}. 
\begin{figure}
\includegraphics[scale=0.65]{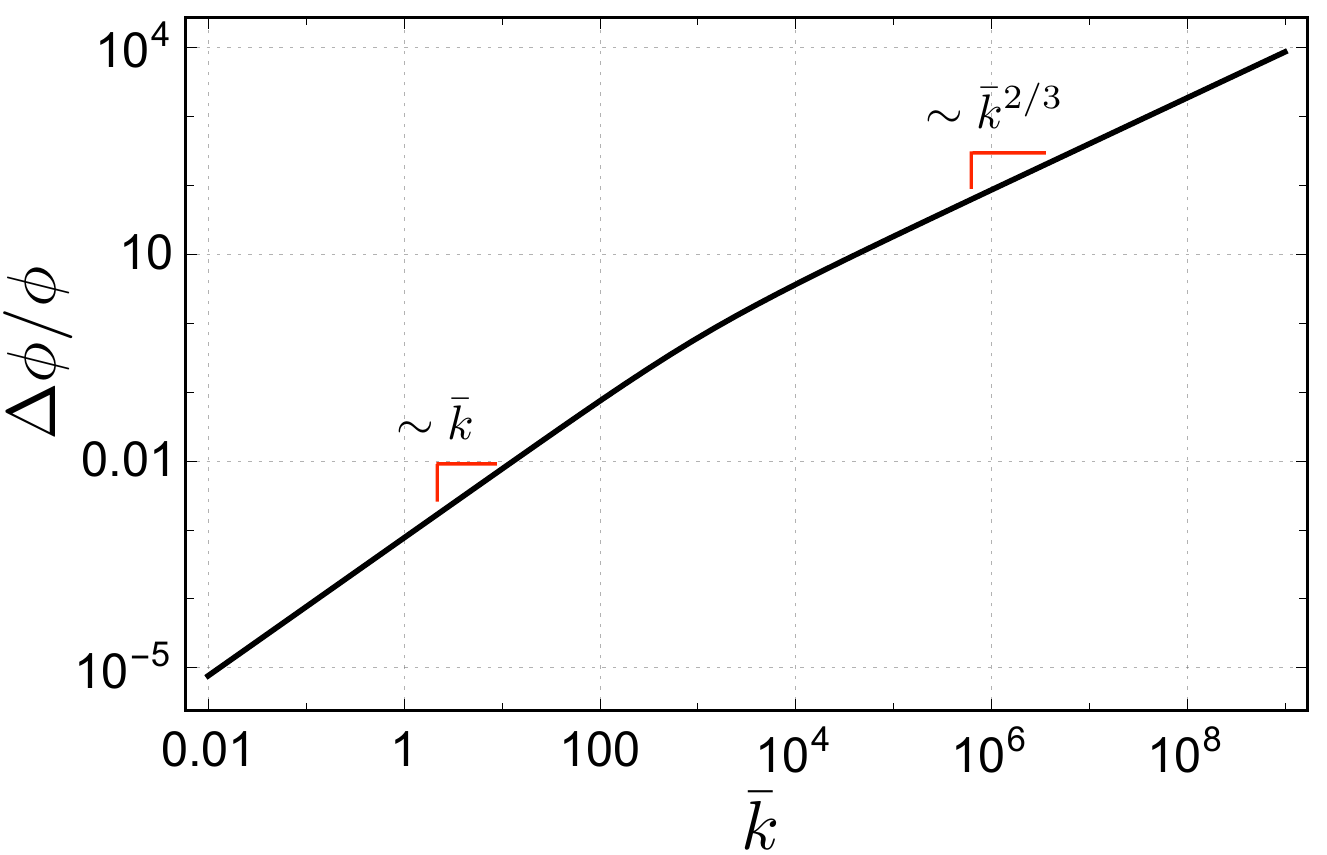}
\caption{(color online) Growth of additive tension renormalization $\Delta \phi$ as a function of the dimensionless spring constant 
$\bar k$. $\phi =100$.  At $k = \tau/\Delta \ell_0$, we can no longer approximate $\Delta \ell$ as being $k$-independent. It decays 
like $ k^{-1/3}$, leading to the shift to $\bar k^{2/3}$ growth in $\Delta \phi$.}
\label{fig: deltaPhiK}
\end{figure}
We leave details of the evaluation of $\lambda_0(t)$ to Appendix~\ref{app: MFT}, and here discuss the results. In the long time limit, 
$\lambda_0(t)$ must approach its equilibrium configuration, a constant $\lambda_0$. Writing $\lambda_0(t) = D k \lambda_0$, 
we find $\lambda_0$ obeys the self-consistent equation (restoring $k_\text{B} T$ for the moment for ease of comparison)
\beq
\label{eq: delta ell equilibrium}
\lambda_0 = \frac{k_\text{B} T}{2} \sum_p \frac{1}{\kappa p^2 + \tau + k \lambda_0},
\eeq
which can be interpreted as a renormalization of the tension $\tau \to \tau + k \langle \Delta \ell \rangle$. In terms of the dimensionless 
tension $\phi$ (see Eq.~\ref{eq: phi def}), this is expressed as the shift $\phi \to \phi+ \Delta \phi$, where $\Delta \phi$ 
satisfies the MFT equation in Eq.~\ref{eq: delta phi}.  The MFT dynamics of a filament attached to a longitudinal spring in 
equilibrium are thus the same as for a semiflexible filament under tension, provided we renormalize tension.

The time-dependent $\lambda_0(t)$ solution is determined by its initial condition. 
We consider the case where the filament is initially in equilibrium with the longitudinal spring, then at $t=0$, we apply a small 
additional tension $\delta f(t)$ to an already tensed filament with tension $\tau \gg \delta f(t)$. In 
Appendix~\ref{app: MFT}, we derive the general solution for this situation. We define the change in the projected length from its equilibrium value,
\beq
\delta \lambda_0(t) =\lambda_0(t)- \lambda_0.
\eeq
At $t=0$, $\delta \lambda_0(t)$ vanishes, and at $t=\infty$ it must plateau to a constant as the system again reaches a new equilibrium. 
The Laplace transform $\delta \lambda_0(z)$ obeys the equation
\beq
\label{eq: lambda0 laplace}
\delta \lambda_0(z) = -\frac{\tilde M(z)}{1+ k \tilde M(z)/z}\delta F(z),
\eeq
where the kernel $\tilde M(z)$ is defined in Eq.~\ref{eq: M tilde}. The function $\delta F(z)$ is the Laplace transform of the time-integrated 
applied tension defined in Eq.~\ref{eq: big F def}. The negative sign arises because putting a filament under tension causes it to extend, 
thereby increasing total projected length, and thus decreasing $\Delta \ell$.

We now examine two cases: 
\beq
\delta f(t) = \left \{ 
\begin{array}{cc}
\text{oscillating:} &  f \sin \omega t  \\ 
\text{constant:} & f \end{array} 
\right. ,
\eeq
corresponding to oscillatory and constant applied tensions respectively. These lead to the Laplace-transformed integrated tensions
\beq
\delta F(z) =
\left \{ 
\begin{array}{cc}
\text{oscillating:} &  f (z/\omega)/(z^2+\omega^2)  \\ 
\text{constant:} & f/z^2 \end{array} 
\right. .
\eeq
The Laplace transform of the MFT longitudinal linear response is trivially related to $\delta \lambda_0(z)$:
\beq
\label{eq: chi lambda0}
\chi_{\Delta \ell}(z) = \delta \lambda_0(z) /f.
\eeq
The remaining step is to take the inverse Laplace transformation in both cases.

We first discuss the oscillatory solution. In the long time limit, only residues corresponding to the purely imaginary poles will remain. The only 
contributing poles are due to $\delta F(z)$, which occur at $z = \pm i \omega$. We can thus substitute $\chi_{\Delta \ell}(z \to - i \omega)$ to 
obtain the long-time oscillatory solution. An alternative derivation is presented later in Sec.~\ref{sec: fluctuations} using the MSRJD formalism. 
Comparing $\tilde M(-i \omega)$ with $\Pi^+(\omega)$ (defined later in Eq.~\ref{eq: pi plus}), and $\chi_{\Delta \ell}(z)$ with the later MSRJD 
result in Eq.~\ref{eq: chi lambda}, we observe that the MFT Langevin equation exactly reproduces the more rigorous MSRJD analysis. 
We thus postulate (but do not prove in this manuscript) that the MFT Langevin equation is capable of providing the exact 
correlations $\langle \Delta \ell(t_1)...\Delta \ell(t_N) \rangle$ for any product of $N$ $\lambda_0(t)$ fields. 

\begin{figure}
\includegraphics[scale=0.62]{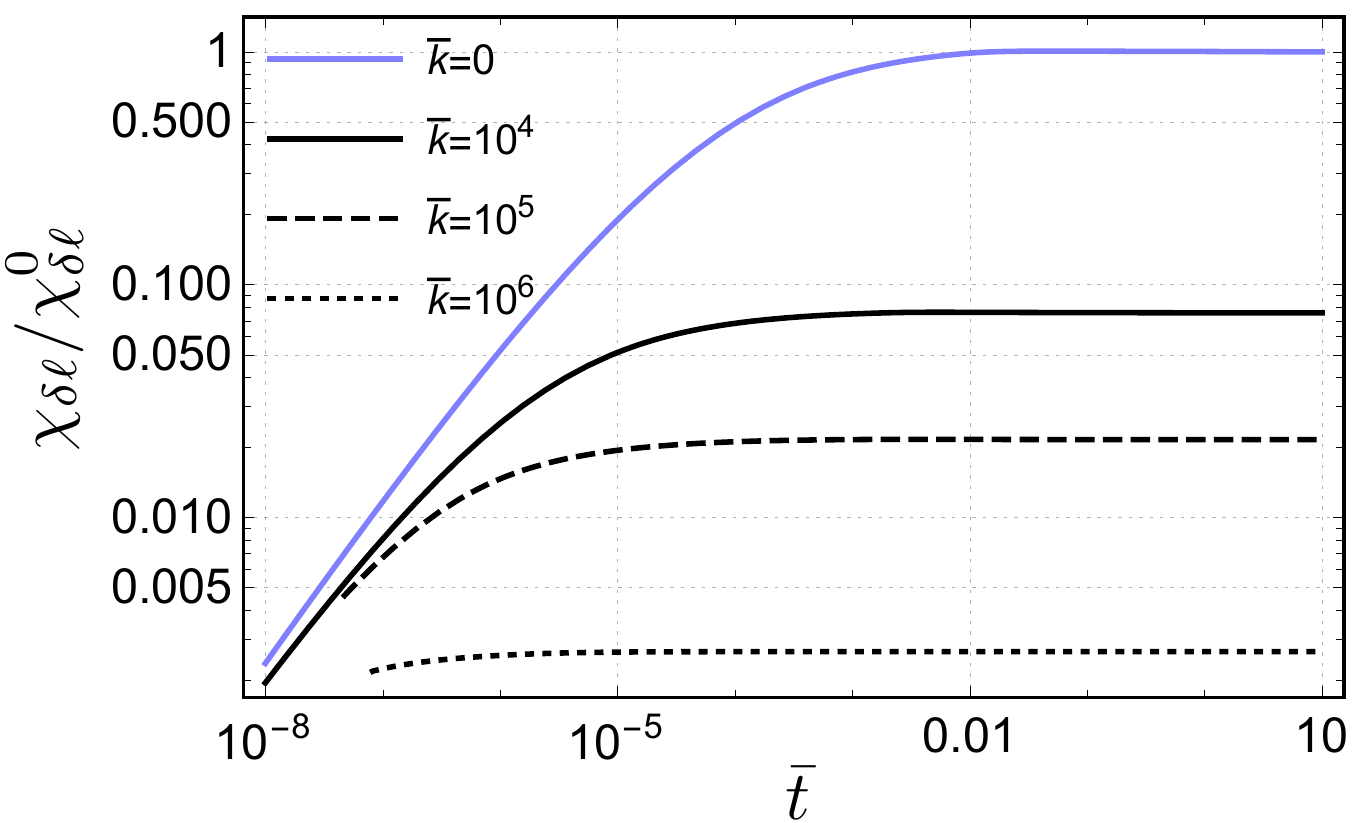}
\caption{(color online) MFT longitudinal linear response normalized by the plateau value $\chi_{\Delta \ell}^0=\chi_{\Delta \ell}(\bar t=\infty,\bar k=0)$ of 
the spring-free filament. $\bar t = t \ell^4 \xi_\perp /\kappa \pi^4$. At early times, there is $\bar t^{3/4}$ growth, but the function does 
not exhibit power-law behavior. The longitudinal spring decreases the relaxation time, roughly proportional to $\bar k^{4/3}$.}
\label{fig: deltaLt}
\end{figure}

The $p$ summation appearing in the kernel $\tilde M(z)$ (Eq.~\ref{eq: M tilde} ) can be performed, but is unwieldy. It is easily performed numerically. 
We used that numerical summation to plot $\delta \langle \Delta \ell \rangle /\delta f$ in Fig.~\ref{fig: deltaLt}. Analytically, we look at the long and 
short time limits, and then comment on the transition between the two. Long/short times correspond to small/large $z$ respectively. 
At long times, $\tilde M(z\to 0) \sim z$, while at short times $\tilde M(z \to \infty) \sim z^{1/4}$. The long-time limit leads to a constant value
 $\lambda_0$, which is determined by the self-consistent Eq.~\ref{eq: delta ell equilibrium} with $\tau$ replaced by $\tau + f$.

At short times, $z$ is large, and so the factor of $k \tilde M(z)/z \sim z^{-3/4}$ is negligible compared to 1. 
We find the simpler expression
\beq
\chi_{\Delta \ell}(z \gg 1) = - \frac{\tilde M(z)}{z^2}.
\eeq
The inverse Laplace transform yields
\beq
\label{eq: spring free chi}
\chi_{\Delta \ell}(t \ll 1) = \frac{k_\text{B} T}{2}\sum_p\frac{e^{-2 D p^2 t \left(\kappa  p^2+\tau \right)}-1}{\left(\kappa  p^2+\tau \right)^2}
\eeq
This is precisely the spring-free result for the longitudinal linear response of a tensed filament.

To extract the short-time behavior, we replace the summation with an integration, extend the limits of integration from $0$ to $\infty$, and make the variable substitution $p \to p (2 D \kappa t)^{1/4}$. At small $t$, the $p^4$ bending terms in the exponent are dominant, leading to
\beq
\chi_{\Delta \ell}(t \ll 1) \approx \frac{k_\text{B}T \ell}{2\pi \kappa^2}(2 D \kappa t)^{3/4} \int_0^\infty \frac{e^{-z^4}-1}{z^4}dz.
\eeq
The integral is $\Gamma(1/4)/3$. From this we find the final result
\beq
\chi_{\Delta \ell}(t \ll 1) \approx \frac{k_\text{B} T \ell  \Gamma(1/4)}{3 \pi 2^{1/4}\kappa^{5/4} \xi_\perp^{3/4}} t^{3/4}.
\eeq
The short time power law growth $t^{3/4}$ is the same as for flexible filaments~\cite{Broedersz2014}. However, this is only the 
leading term at short time. Due to the presence of $\tau$, the filament breaks self-similarity and the function does not obey a power law.

The short-time longitudinal response is bending dominated, and independent of the spring. 
From Eq.~\ref{eq: lambda0 laplace}, we expect the longitudinal spring to become important when $k \tilde M(z) /z >1$. 
As $k\to \infty$, the $\tilde M(z)/z$ in the numerator and denominator cancel out, leaving the inverse Laplace transform of $-f/z k$, which 
gives a constant. Thus, the spring shortens the relaxation time.  Since, in the short time limit $\tilde M(z) \sim z^{1/4}$, this suggests that the relaxation 
time to equilibrium decreases with increasing spring constant like $k^{-4/3}$. 

In Fig.~\ref{fig: deltaLt}, we plot the response function by performing a numerical inverse Laplace transform of Eq.~\ref{eq: chi lambda0}  
using Eq.~\ref{eq: M tilde phi}, for several values of $k$. It exhibits the predicted $t^{3/4}$ spring-free growth. Increasing $k$ shortens the equilibration time.

To conclude the section, we consider how the decay rates of normal modes are altered in the MFT. From 
Eq.~\ref{eq: up nonavg}, specifying $u_p(0)$ then averaging over the noise suggests that normal 
modes obey a time-dependent decay rate, $\tau_\text{decay}$, given by
\beq
\tau_\text{decay}^{-1}(t) = \xi_\perp^{-1} \left(\kappa p^4 + \tau p^2 + k t^{-1} \int_0^t \lambda_0(t') dt' \right).
\eeq
At short times, $\lambda_0(t) \sim t^{3/4}$, which implies an additional stretched exponential prefactor $ \langle u_p(t) \rangle \sim e^{- k t^{7/4}/\xi_\perp}$ 
(again, the average is over noise and $u_p(0)$ is specified). Since at small times $t> t^{7/4}$, we expect this effect to be 
difficult to observe in experiment.

\subsection{Fluctuations/Random phase approximation}
\label{sec: fluctuations}
The saddle-point approximation, while accurately calculating $\langle \Delta \ell(t) \rangle$, does not address multipoint correlations of $\Delta \ell(t)$.
This prevents us from understanding how the spring-induced nonlinearity affects dynamic fluctuations of $\Delta \ell(t)$.  We define the 
longitudinal correlator
\beq
\label{eq: long c def}
C_{\delta \ell}(t,t') = \langle \Delta \ell(t) \Delta \ell(t') \rangle - \langle \Delta \ell(t) \rangle \langle \Delta \ell(t') \rangle
\eeq
to be the correlation functions of the end-to-end distance.  
This quantity is related to the dynamic shear modulus~\cite{Gittes1997,gittes1998G} and informs 
frequency-dependent activity microscopy~\cite{Lissek2018,kernes2020equilibrium}. 

We account for fluctuations by expanding the trace-log term $\text{Tr}\ln {\bf G}^{-1}(\lambda_0 + \delta \lambda, \delta \bar \lambda)$ of the 
action (Eq.~\ref{eq: A lambda}) in powers of $\delta \lambda,\,\delta \bar \lambda$, about the saddle-point. In principle, one may carry out the 
expansion to arbitrary order. We stop at the quadratic terms. This truncation is a valid approximation for stiff filaments, where 
the equilibrium end-to-end contraction is small compared to contour length.

Since we are considering fluctuations about equilibrium, time-translation invariance allows us to Fourier transform to the frequency domain. 
In frequency/wavenumber space, the propagators and correlators appearing in the expansion refer to the saddle-point/MFT values:
\begin{subequations}
\beq
\label{eq: G MFT}
\bar G_p^+(\omega) = \frac{2/\ell}{-i \omega + Dp^2 ( \kappa p^2 + \tau + k \lambda_0)},
\eeq
\beq
\label{eq: C MFT}
\bar C_p(\omega) = \frac{4D/\ell}{\omega^2 + [Dp^2 ( \kappa p^2 + \tau + k \lambda_0)]^2}.
\eeq
\end{subequations}

In Appendix~\ref{app: polarization}, we carry out the trace-log expansion to quadratic order, yielding the 
Gaussian approximation to the action at the saddle point:
\beq
\label{eq: effective action}
\mathcal{A}_\text{eff}[\delta \bar \lambda, \delta \lambda] = \frac{1}{2} \int\frac{ d\omega}{2\pi} \delta \bm{\lambda}^T_{\omega} \mathbf{M}^{-1}_\omega \delta{\bm \lambda}_{-\omega}, 
\eeq
where the matrix $\mathbf{M}^{-1}_\omega$ is defined in Eq.~\ref{eq: M inverse}. This is our final expression 
for the effective action $\mathcal{A}_\text{eff}$. We are primarily concerned with the inverse $\mathbf{M}_\omega$. 
It is related to fluctuations in the projected length, and its linear response to applied tension. We compute
\beq
\label{eq: M}
\mathbf{M}_\omega = \left( \begin{array}{cc} 0 & \frac{1}{1+ k \Pi^-_\omega} \\  \frac{1}{1+ k \Pi^+_\omega} & \frac{ \Pi^0_\omega}{|1+k \Pi^+_\omega|^2} \end{array} \right),
\eeq
where the polarization functions, $\Pi_\omega^{\pm}$ and $\Pi_\omega^{0}$, are defined as
\beq
\label{eq: pi plus}
\Pi^\pm(\omega) =  \sum_p \frac{D p^4}{\gamma_p (\mp i \omega+ 2 D \gamma_p)},
\eeq
\beq
\label{eq: pi 0}
\Pi^0(\omega) = \sum_p \frac{2 D p^4}{\gamma_p (\omega^2 + 4 D^2 \gamma_p^2)}.
\eeq
The $\Pi^0$ function is precisely the Fourier transform of the spring-free correlator~\cite{gittes1998G}. The $\pm$ 
functions are complex conjugates of one another, {\em i.e.}~$\Pi^+ = (\Pi^-)^*$. They can be related to $\Pi^0$ via a 
fluctuation-dissipation-like relation (Eq.~\ref{eq: pi FDT}).
Using Eq.~\ref{eq: chi lambda}, we relate the $\Pi^+$ function to the longitudinal linear response via 
\beq
\label{eq: chi long}
\chi_{\Delta \ell}(\omega) = \frac{\Pi^+(\omega)}{1+ k \Pi^+(\omega)}.
\eeq
Comparison of the correlation function with $\chi_{\Delta \ell}$ confirms that the fluctuation-dissipation theorem is satisfied. 

\begin{figure}
\includegraphics[scale=0.6]{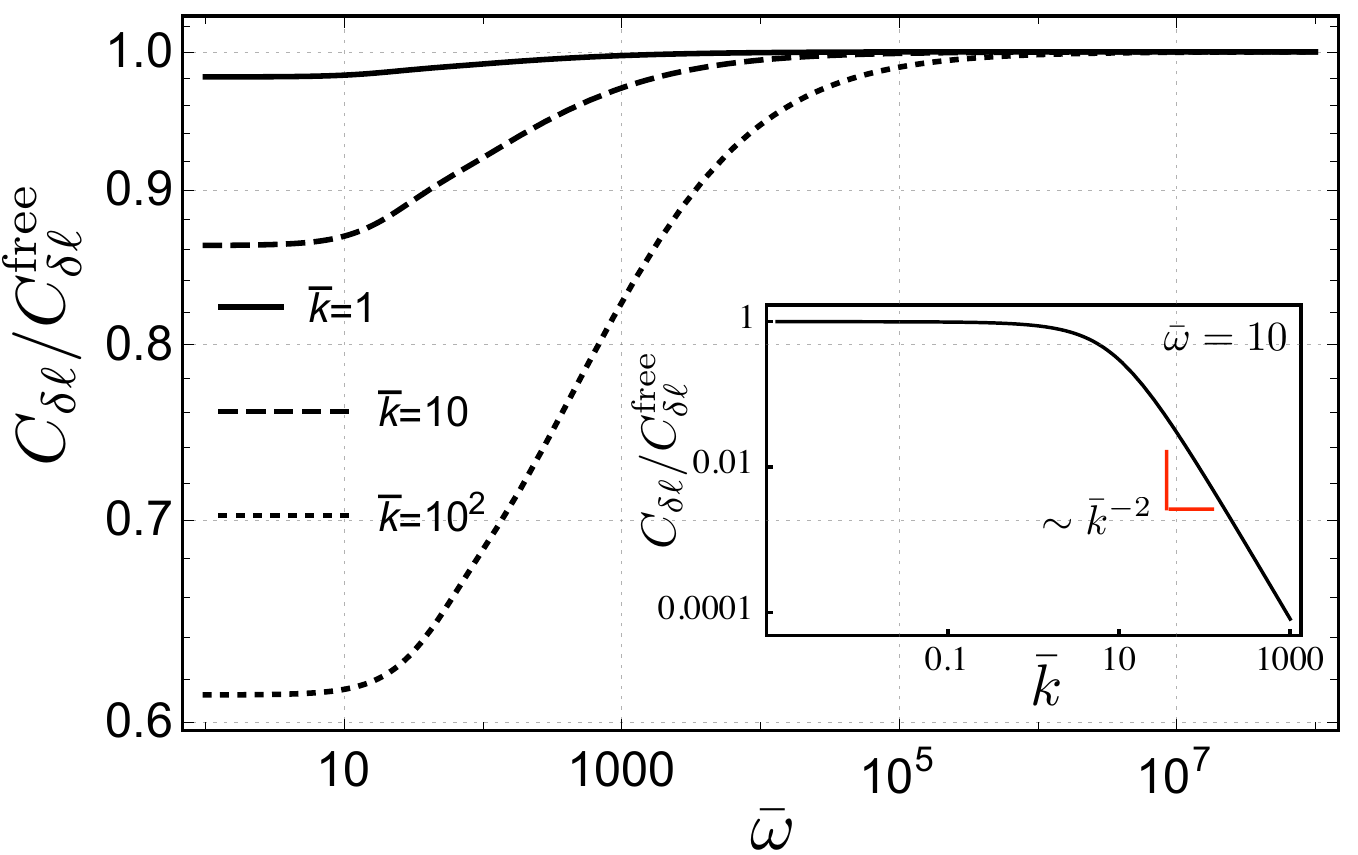}
\caption{(color online) Ratio of the longitudinal correlator to its spring-free value. $\phi =1$. At high frequency, individual modes have not relaxed 
to a new equilibrium that accounts for the longitudinal spring, so the ratio flattens to one. As frequency decreases, we approach the 
static result of Eq.~\ref{eq: delta ell equilibrium}, whereby we find a reduced amplitude, with zero slope. The inset shows that, for a fixed 
frequency ($\bar \omega=10$) and zero tension $\phi=0$, the ratio decays as $\bar k^{-2}$ after passing a frequency-dependent cross-over spring constant 
$k^*$.}
\label{fig: Clong}
\end{figure}

The ratio
\beq
\label{eq: lambda fluctuations}
\frac{C_{\delta \ell}(\omega)}{C_{\delta \ell}^\text{free}(\omega)} = \frac{1}{|1+  k \Pi^+(\omega)|^2},
\eeq
of projected length fluctuations in the presence/absence of a longitudinal spring, makes the effect of the spring more transparent. That ratio is 
plotted in Fig.~\ref{fig: Clong}.
We first analyze the $k$ dependence of the ratio. If we slowly increase $k$, we see that, below $|k \Pi^+(\omega)| =1$, 
there is little deviation from the spring-free result. When $k$ is large enough to exceed the bound $|k \Pi^+(\omega)| =1$, then 
the end-to-end fluctuations diminish as $k^{-2}$. This is supported numerically -- see the inset of Fig.~\ref{fig: Clong}. It is interesting that 
below a certain value of $k$, the effect of the longitudinal spring on the end-to-end distance fluctuations is {\em screened}.  This shows that
the effect of the longitudinal spring goes beyond tension renormalization.  When looking at the dynamics of the end-to-end fluctuations, we now 
observe the filament length stored in the various normal modes at different times interact (through the spring) to make the dynamics of the
end-to-end length more complex.  

The value $k^*$ beyond which screening breaks down is, itself, frequency dependent. Specifically, $k^* \approx |\Pi^+(\omega)|^{-1}$. 
Since $\Pi^+(\omega)$ is decreasing with $\omega$, screening breaks down at smaller values $k^*$ as $\omega$ decreases, bottoming out in the 
static limit ($\omega =0$) with a minimum value $\bar k_\text{min}^* = \left[\sum_{n=1}^\infty (n^2 + \phi)^{-2}\right]^{-1}$.  Below this spring constant 
$\bar k_\text{min}^*$, screening occurs at all frequencies.

The frequency dependence of the ratio of the correlators with and without the spring can be understood similarly. At $\omega=0$, if $k> k_\text{min}^*$, 
then the longitudinal spring shifts the longitudinal correlator to its mean-field result. In the opposite limit where $\omega \to \infty$, screening becomes 
perfectly effective, and there is no deviation from the spring-free result. In the main panel of Fig.~\ref{fig: Clong}, we see that the longitudinal correlator 
transitions from the spring-dominant, mean-field result to the spring-free result across a range of frequencies that increases with $k$.

By examining the saddle point analysis, we obtain further insight into which of the perturbative corrections we have taken into account in 
this approach. Examining the action in Eq.~\ref{eq: effective action}, we claim that it is a renormalization of the dashed line 
propagators of the original $(\bar u, u)$ theory. Since all dashed-line renormalizations are necessarily bubble type diagrams, $\mathbf{M}$ 
contains the contributions from all two-bubbles (the general $n$-bubble subdiagram is a solid line loop with exactly $n$ outgoing dashed lines). 
Taking higher order terms in the expansion of the trace-log will result in bubbles with $n>2$ external dashed lines, which are exactly the $n$-bubbles. 
The fluctuation expansion is not just a $k$ expansion, but a systematic inclusion of higher number bubbles.  

We can estimate the relative importance of successive terms. The one-bubble returns just the static change in projected length $\Delta \ell_0$. The 
two-bubble $\sim \sum_p G_p^2(\omega=0) \sim \p_\tau \langle \Delta \ell_0 \rangle$, is proportional to the static susceptibility $\chi_{\Delta \ell}$, 
with each higher order gaining another derivative of the projected length with respect to 
$\p_\tau$. Since each derivative lowers the summand by $p^{-2}$, successive terms quickly become small. 

Classifying the diagrams in Fig.~\ref{fig: Ok2} of the perturbation series, to $\mathcal{O}(k^2)$ they can be divided into one-bubbles, two-bubbles, and the rest. 
Per the saddle-point analysis, the one/two-bubbles are the leading/subleading terms, corresponding  to renormalization of the effective tension/spring constant. 
This is consistent with our analysis in Sec.~\ref{sec: perturbation theory}, where our grouping of diagrams into dominant and 
subdominant classes was in fact a grouping into $n$-bubbles.

Returning to our analysis of the effective action, we observe that the resummation of bubble diagrams is an approximation known as the 
{\em random phase approximation} (RPA)~\cite{altland2010}. The RPA applies only to the dashed line, which in any actual diagram must be attached to 
two solid lines according to the rules in Fig.~\ref{fig: lambda u u}. The two-bubble renormalized vertex is given by the 
diagrams in Fig.~\ref{fig: RPA}, which yield the equations:
\begin{figure}
\includegraphics[scale=0.4]{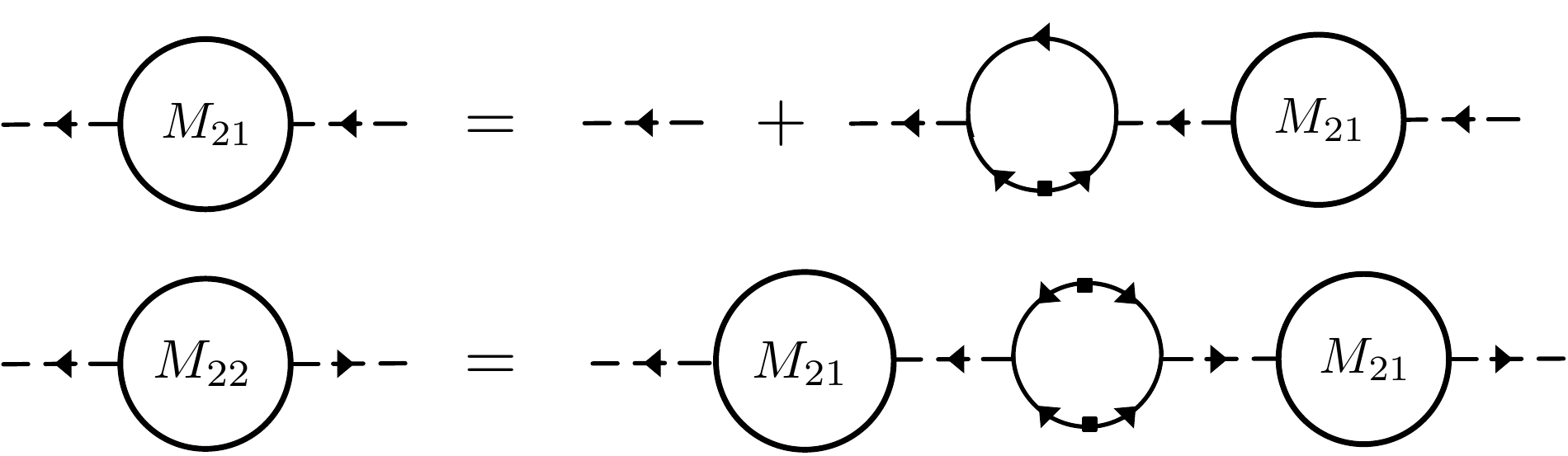}
\caption{Random phase approximation for computing renormalized interaction vertex. When used in a diagram, the 
directed dashed lines must join to external solid lines according to Fig.~\ref{fig: lambda u u}. $M_{ij}$ refers to the matrix elements of $\mathbf{M}$. 
$M_{21} = \langle \lambda \bar \lambda \rangle$ is directed from a vertex with two incoming lines, to one with an incoming and outgoing.}
\label{fig: RPA}
\end{figure}
\begin{subequations}
\beq
M_{21}(\omega) = 1 - \frac{kD\ell^2}{4}\sum_p \int \frac{d\omega'}{2\pi} p^4 \bar C_p(\omega'-\omega) \bar G_p^+(\omega'),
\eeq
\beq
M_{22}(\omega) = |M_{21}(\omega)|^2  \sum_p\int \frac{d\omega'}{2\pi} \frac{p^4 \ell^2}{4}\bar C_p(\omega-\omega') \bar C_p(\omega').
\eeq
\end{subequations}
Solving these reproduces Eq.~\ref{eq: M}, thus confirming our claim. Since the dashed lines appear only in combination with $k$, the 
RPA amounts to a renormalization of $k$. $M_{21}$ and $M_{22}$ represent effective vertices, whose lowest order terms 
reproduce diagrams B1 and B2, and B3 respectively in Fig.~\ref{fig: Ok2}.  
\begin{figure}
\includegraphics[scale=0.6]{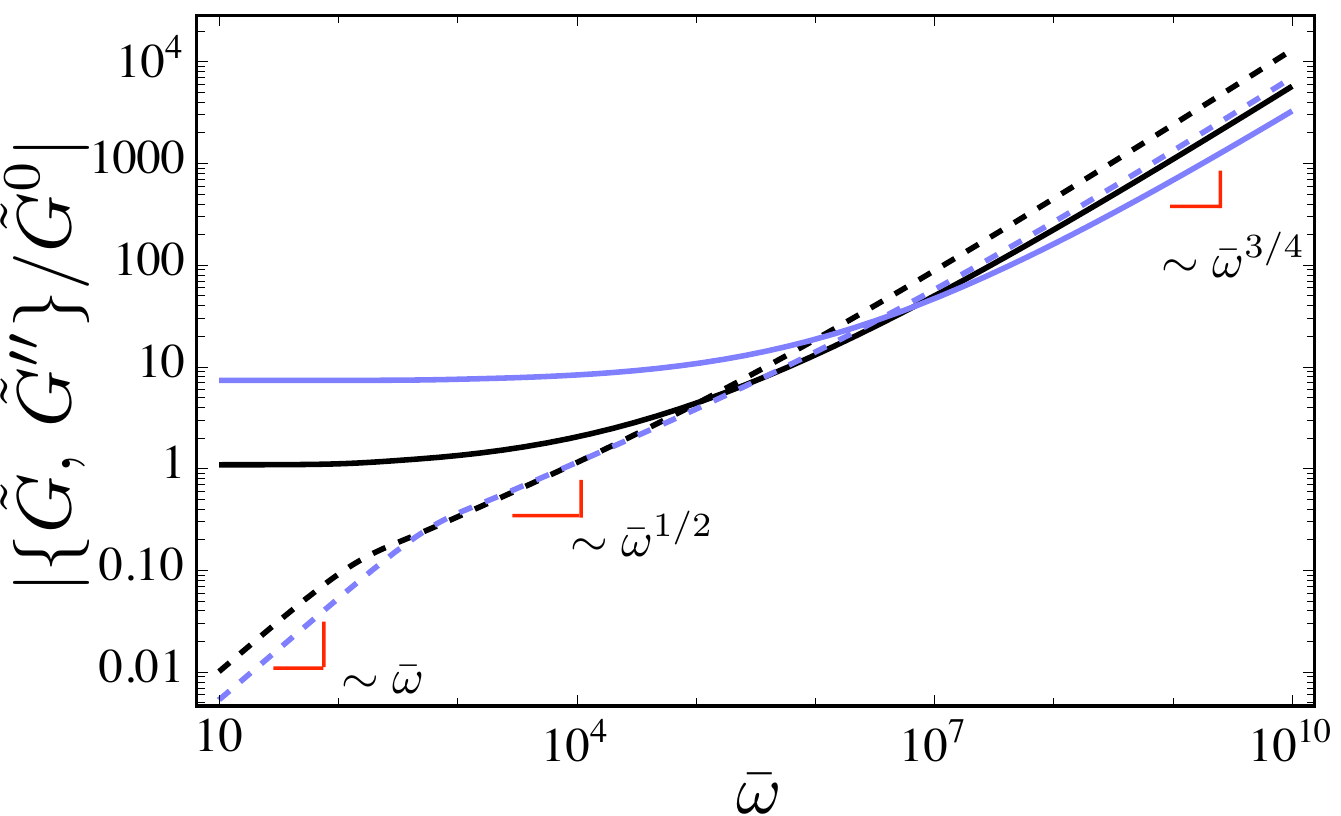}
\includegraphics[scale=0.6]{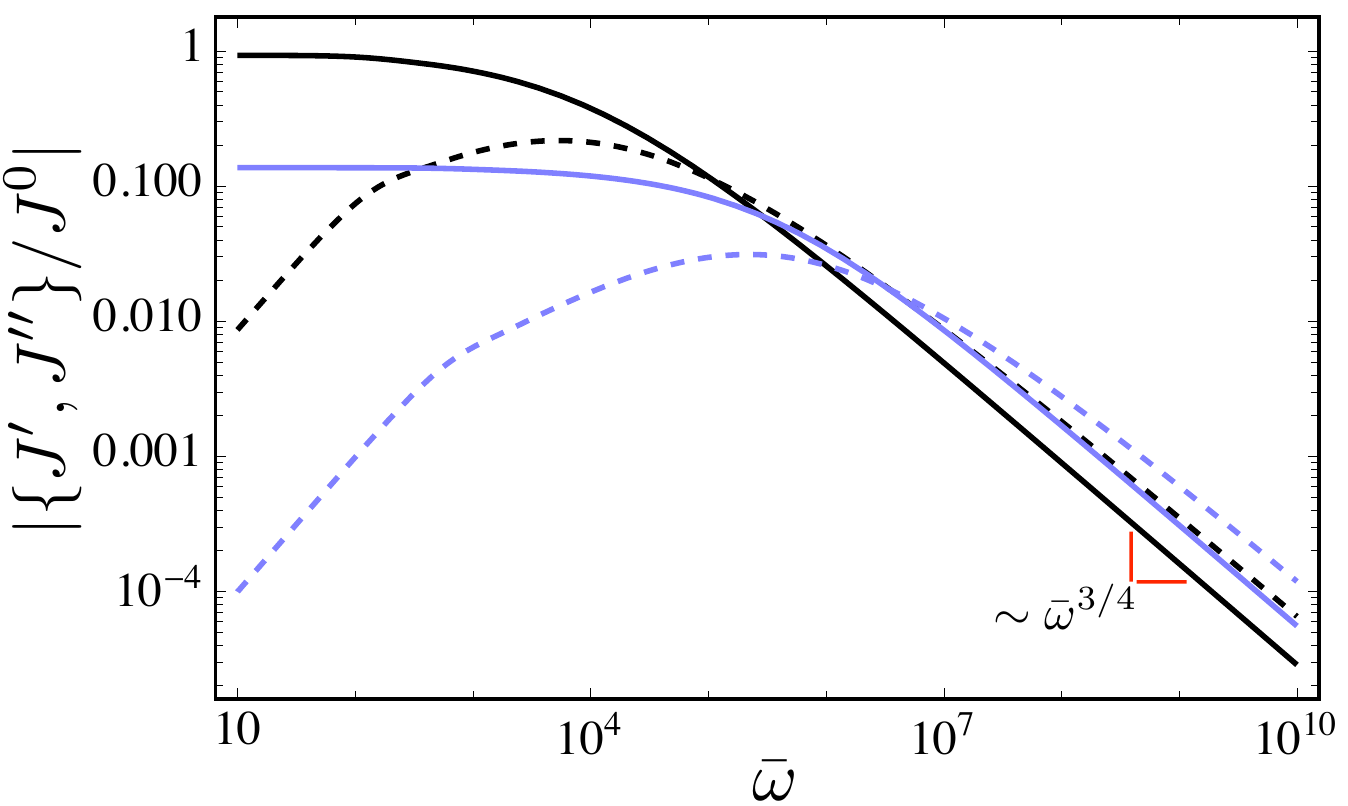}
\caption{(color online) Real (solid lines) and imaginary (dashed lines) parts of the shear modulus $\tilde{G}(\omega)$ and network 
compliance $J(\omega)$ for $\bar k=0$ (black) and for $\bar k=10^4$ (blue). $\phi=10^2$.  $\tilde{G}$ and $J$ are normalized by their spring-free 
plateau values.  The transition of $\tilde{G}(\omega)$ from $\omega^{1/2}$ to $\omega^{3/4}$ scaling signals the shift from tension- 
to bending-dominated behavior~\cite{seifert1996,granek1997R}. The longitudinal spring does not alter the power-law 
dependence, but shifts the cross-over between them to higher frequencies.}
\label{fig: J and G}
\end{figure}

Finally, we consider the longitudinal response function given in Eq.~\ref{eq: chi long}. This provides an estimate for the high-frequency behavior of the dynamic shear modulus of semiflexible networks~\cite{hiraiwa2008,obermayer2009tension}, via the 
relation $\tilde{G}(\omega) = \frac{1}{15} \rho \ell \chi_\omega^{-1} - i \omega \eta$~\cite{gittes1998G}, where $\rho$ denotes the density of filaments. 
Note that the dynamic shear modulus, $\tilde{G}(\omega)$, must be distinguished from our earlier definitions of propagators. 
Ignoring the viscous term, in Fig.~\ref{fig: J and G} we plot both $J(\omega)$ and $\tilde{G}(\omega)$ for both $\bar k=0$ and 
$\bar k=10^4$. $\tilde G(\omega)$ possesses three distinct scaling regimes, regardless of the longitudinal spring: a low-frequency regime 
$\sim  \omega$, an intermediate-frequency regime $\sim  \omega^{1/2}$, and a high-frequency bending regime $\sim  \omega^{3/4}$. The 
spring does not affect this scaling, but shifts the transition region to higher $\omega$ as $k$ increases, which is 
consistent with our assertion that tension renormalization is the spring's main effect.

\section{conclusion}
\label{sec: conclusion}
We have extended the static analysis of Ref.~\cite{kernes2020equilibrium} to include the dynamics of the fluctuations of a filament in network, 
whose linear compliance is modeled as a hookean spring attached to the boundary. The principal motive behind these calculations is to provide a precise prediction for the dynamical fluctuations of filaments based on $\kappa$, $\tau$, and $k$, that can then be used to perform local activity microscopy.

The addition of the spring boundary condition introduces a nonlinearity into the problem, which is peculiar in the sense that it is nonlocal in space but
local in time.  It depends at each instant on the projected length of the whole filament. The peculiarity stems from our 
assumption of instantaneous tension propagation. The strength of the nonlinearity can be externally governed via the spring constant $k$. 

For a filament bound to a larger network, the spring constant $k$ approximates the compliance of the entire surrounding network. To get an estimate for experimentally relevant values of parameters, we use as an example an F-actin network with shear modulus $G \approx 100 \text{Pa}$, and 
mesh spacing  $\xi \approx 0.5 \mu\text{m}$, which we assume is comparable to the mean distance between consecutive cross links along the 
same filament. Using the relation $G \sim k/\xi$~\cite{mackintosh1995}, we estimate a spring constant of $0.05 \text{pN/nm}$~\cite{kernes2020equilibrium}. 
We further assume a persistence length $\ell_p = \kappa/k_\text{B} T$ that is 
approximately an order of magnitude greater than the filament segment 
length, and $k_\text{B} T \approx 4 \,\text{pN nm}$. These suggest $\bar k \sim 10^3$. In these dimensionless units, a tension of 
1pN corresponds to $\phi \sim 10^2$.

The most direct and quantitatively precise experimental test of this analysis is directly examining the dynamics of a single 
filament tethered to a bead in an optical (or magnetic) trap. In that case, one can independently control both the mean tension in the filament and the effective
spring constant $k$ by varying the position of the optical trap and its intensity respectively.  In this setup, one may imagine two distinct types of 
measurements. One could observe the end-to-end length fluctuations by tracking the bead in the trap with high spatial and temporal resolution. 
Or, one could observe the undulations of the filament directly, which would allow one to measure the $u$ correlation functions computed here. 
In both of these cases, one might also measure the response functions by observing the response of either $u$ or the end-to-end distance of the filament to changes in the trap's center.

We find the main effect of the spring is to renormalize tension. Even for untensed filaments, once a spring is added, the filament behaves
as if it were under tension $\tau_\text{R} \approx k \langle \Delta \ell \rangle_0$. Perhaps, this 
blending of spring effects into an effective tension explains the success of 
previous theories, which have neglected nonlinearities introduced by a longitudinal boundary spring~\cite{mackintosh1995,Broedersz2014}. 
We have shown that an increasing spring constant decreases the relaxation times of all the fluctuating degrees of freedom of the filament. 
Given a fixed external tension $\tau$, there is a scale $k^* \approx 12\ \kappa \tau /k_\text{B}T \ell^2$ of the external spring constant, above which 
the dominant contribution to tension comes from the spring and not the bare applied tension. Using this crossover, we estimate the minimum 
spring constant whose effect on filament dynamics should be 
observable. For typical filaments on the order of microns (with 
persistence lengths greater than their contour length) and tensions $\sim 10$pN, we expect this transition to occur at $k \approx 1$(pN/nm). 
This is achievable near the upper limit of optical trap strength ($\sim100$pN/$100$nm), or by alternatively using a 
magnetic trap that can achieve higher $k$ values.

One could alternatively use optical tweezers to exert localized forces within a network of filaments, putting some of them under 
tension $\tau$. We predict that increasing $\tau$ will not affect fluctuations up until a transition tension 
$\tau^* \approx k k_\text{B} T \ell^2/ 12 \kappa$, after which the amplitude of fluctuations will decrease as $\tau^{-2}$. In the absence of the 
network compliance (the spring in our model), the transition occurs at a lower tension, which is frequency-dependent.  One may also look for 
nontrivial changes in filament tension and fluctuations as a function of network stiffness. Since the change in 
tension due to the spring is $\Delta \tau \sim k\Delta \ell$, for 
sufficiently stiff networks where the effective spring constant $k>k^*$, the change in tension switches from a linear $k$ dependence to a weaker one 
$\sim k^{2/3}$, due to the shortening of $\Delta \ell$. This leads to a $k^{-4/3}$ decrease in the amplitude of transverse fluctuations. 

We also considered fluctuations of the end-to-end projected length of the filament, and its response to an applied tension. We found that including the 
external longitudinal spring does not affect the short-time longitudinal, linear response of projected length to an abrupt change in 
applied tension.  The change in projected length grows initially like $t^{3/4}$, but does not exhibit a power law at longer times. The longitudinal spring, 
does, however,  shortening the relaxation time of the end-to-end length by a factor $\sim k^{-4/3}$. From the response function of the 
end-to-end distance of a single filament to oscillatory forcing, we can predict the collective dynamic shear modulus of the network using now standard 
arguments.  We find that the spring shifts the transition from tension dominated, $G(\omega) \sim \omega^{1/2}$, to bending dominated, $G(\omega) \sim \omega^{3/4}$, to higher frequencies. 

Finally, there is an additional frequency-dependent effect that can be observed from fluctuations in the end-to-end projected length, which arises as a 
result of the nonlinear interaction the spring induces on normal modes. In the static, $\omega \to 0$ limit, the amplitude of end-to-end fluctuations
will be lower than that for a filament not attached to a longitudinal boundary spring. As frequency increases, however, the effect of the 
spring diminishes, approaching the spring-free result as $\omega \to \infty$. At high frequencies the normal modes adjust so as to 
screen the effect of the longitudinal spring. We report a minimum value $\bar k_\text{min}^*$, below which the longitudinal spring is 
screened at all frequencies.  This occurs when $\bar k_\text{min}^* = \left[\sum_n (n^2 + \phi)^{-2}\right]^{-1}$.  This minimal spring stiffness 
necessary for complete screening grows with applied tension as $\sim \tau^{3/2}$.

Future directions for this work include a first-principles calculation of the effective spring constant $k$ representing the network.  At least, 
one may imagine pursuing a type of self-consistent analysis by demanding that the force extension relation of the filament coupled to the spring is 
identical to those of the network filaments, whose collective elasticity is represented by that spring.  Secondly,  one may consider how the transverse 
undulations of a filament in the network (represented by external springs coupled to the end of that filament) behave in response to nonequilibrium 
driving, such as would be experienced by the filament in a network driven by endogenous molecular motors. 

\section{acknowledgements}
The authors would like to thank the Botvinick group at UC Irvine for experiments that motivated the study of this problem. 
The authors acknowledge partial support from NSF-DMR-1709785.
\appendix

\section{Diagrammatic perturbation theory to $\mathcal{O}(k^2)$}
\label{app: sigma}
We compute the adjusted self energy given in Eq.~\ref{eq: sigma tilde} to $\mathcal{O}(k^2)$. For readability, in this section we drop the 
$0$ superscript, with $\gamma_p$ referring to $\gamma_p^0$. When we refer to diagrams appearing in Fig.~\ref{fig: Ok2}, we are including 
not only the diagram, but also its combinatorial factor for contracting the legs.  We also include a factor of $(- Dk \ell^2/8)^n/n!$ at $\mathcal{O}(k^n)$. 
Diagram B1 has combinatorial factor $2$, A1 has $1$, all C and D diagrams have $2^3$, B2 has $2^2$, B3 has $2^3$, and A2 has $2^2$.

We calculate for the $\mathcal{O}(k)$ diagrams:
\beq
A1= -\frac{1}{2} k p^2 \sum_q \frac{q^2}{\gamma_{q}},
\eeq
and
\beq
B1 = - k \frac{p^4}{\gamma_p}.
\eeq
For the D diagrams, $D1$ and $D2$ vanish due to a closed response loop. D3 and D4 give identical contributions, 
leading us to write the D contribution
\beq
D3+D4= \frac{ k^2 p^8}{\gamma_p^3}.
\eeq
All three of the C diagrams give the same contribution. Summing these gives
\beq
C1+C2+C3 = k^2\frac{3 D p^8}{\gamma_p^2(-i \omega + 3 D \gamma_p)}.
\eeq
Lastly, for the B diagrams,
\beq
B2 = k^2\frac{p^4}{\gamma_p}\sum_{q}\frac{Dq^4}{\gamma_q (-i \omega+ 2D\gamma_q + D \gamma_p)},
\eeq
and
\beq
B3 = k^2 p^4 \sum_{q}\frac{Dq^4}{2\gamma_q^2 (-i \omega+ 2D\gamma_q + D \gamma_p)}.
\eeq
Taking the sum, we simplify to
\beq
B2+B3 = \frac{k^2 p^4}{2 \gamma_p} \sum_q \frac{q^4}{\gamma_q^2}\left( 1 - \frac{- i \omega}{- i \omega + D(2\gamma_q + \gamma_p)}\right).
\eeq
The last diagram is
\beq
A2 = k^2 p^2 \left( \sum_q \frac{q^2}{2\gamma_q}\right)\left( \sum_q \frac{q^4}{2\gamma_q^2}\right).
\eeq
The adjusted self-energy is simply the sum of these contributions. Altogether we find
\begin{widetext}
\begin{eqnarray}
\tilde \Sigma_p(\omega) &=& -k \left( \frac{p^4}{\gamma_p} + \frac{p^2}{2}\sum_q \frac{q^2}{\gamma_q} \right)+ k^2 \bigg[ \frac{p^8}{\gamma_p^2} \left( \frac{1}{\gamma_p} + \frac{3 D}{-i \omega + 3 D \gamma_p} \right)
+\frac{p^4}{2 \gamma_p} \sum_q \frac{q^4}{\gamma_q^2}\left(1 - \frac{- i \omega}{- i \omega + D(2\gamma_q + \gamma_p)} \right) 
 \nonumber \\
&+&\frac{1}{4}p^2 \left( \sum_q \frac{q^2}{\gamma_q}\right)\left( \sum_q \frac{q^4}{\gamma_q^2}\right) \bigg] + \mathcal{O}(k^3).
\end{eqnarray}

We may rewrite this in terms of the dimensionless tension, $\phi = \frac{\tau \ell^2}{\pi^2 \kappa}$, and dimensionless frequency,
$\bar \omega = \omega \ell^4/(D \kappa \pi^4)$, as 
\begin{eqnarray}
\label{eq: sigma phi}
\tilde \Sigma_n(\omega) &=& - \frac{k k_\text{B} T}{\kappa} \left( \frac{n^2}{n^2 + \phi} + \frac{n^2}{2} \sum_m \frac{1}{m^2+\phi} \right) + \frac{k^2 \ell^4 k_\text{B}^2T^2}{\kappa^3 \pi^4} \bigg[ \frac{n^2}{(n^2 +\phi)^3} + \frac{3 n^4}{-i \bar \omega + 3 n^2(n^2+\phi)} 
\nonumber \\
&+& \frac{1}{2}\frac{n^2}{n^2+\phi}\sum_m \frac{1}{(m^2+\phi)^2}\left(1 - \frac{-i \bar \omega}{-i \bar \omega + 2 m^2 (m^2+\phi) + n^2(n^2+\phi)}\right) + \frac{n^2}{4}\sum_{m,m'} \frac{1}{(m^2+\phi)(m'^2+\phi)^2} \bigg],
\end{eqnarray}
where $m,\,m',$ and $n$ are positive integers, and we have restored factors of $k_\text{B}T$.
\end{widetext}
Rewriting $\tilde \Sigma_p(\omega)$ in terms of the dimensionless wavenumber $\bar p = p \sqrt{\kappa/\tau}$ instead leads to 
Eq.~\ref{eq: sigma o k2}. We can categorize several of the diagrams in terms of the $n$-bubble expansion. 
Diagrams of type A contain one-bubbles, and generate a shift in the effective tension. Diagrams of type B contain
 two-bubbles, and generate a shift in the effective spring constant $k$. The remaining diagrams are single line topologies.

\section{Mean-field theory solution in time-domain}
\label{app: MFT}
Our starting point is Eq.~\ref{eq: MFT equations}.  We begin by defining the integrated projected length 
\beq
\Lambda(t) = \int^t \lambda_0(t') dt',
\eeq
as the antiderivative of $\lambda_0(t)$. The differential equations of motion (Eq.~\ref{eq: MFT eq1}) governing the 
normal modes can be solved in terms of $\lambda_0(t)$. We find
\beq
\label{eq: up nonavg}
u_p(t) = u_p(0) \tilde \chi_p(t,0) + \int_0^t dt' \tilde \chi_p(t,t') \zeta_p(t'),
\eeq
where we defined
\beq
\label{eq: chi app1}
\tilde \chi_q(t,t') = e^{-D(\kappa q^4+ \tau q^2) (t-t') - D k q^2 (\Lambda(t) - \Lambda(t') )},
\eeq
in agreement with the notation of Refs.~\cite{hiraiwa2008,obermayer2009tension}. The initial condition 
$u_p(0)$ may either be specified, or treated as a random variable. Using Eq.~\ref{eq: MFT eq2}, we can eliminate the 
normal modes in favor of a single PIDE governing $\lambda_0(t)$. We obtain
\beq
\label{eq: initial PIDE} 
\frac{d \Lambda}{dt} = \frac{\ell}{4} \sum_q q^2 \left\{ \chi_q^2(t,0) \langle u_q^2(0) \rangle + \frac{4 D}{\ell} \int_0^t \chi_q^2(t,t') dt' \right\}.
\eeq
The brackets around $u_p(0)$ indicate an average over these initial amplitudes. Since the average over the initial amplitudes ($u_q^2$) 
may be taken with respect to any ensemble, this equation can describe the relaxation of a 
nonequilibrium state. In this manuscript however, we will be concerned with the case where $u_p(0)$ is sampled from the equilibrium ensemble.

We begin our analysis with the long-time or equilibrium limit. We implement the long-time limit 
by removing the initial condition and setting the lower limit of integration to $-\infty$. This gives the 
long-time limit PIDE
\beq
\label{eq: PIDE sum}
\frac{d \Lambda}{dt} = D \ell \sum_p p^2 \int_{-\infty}^t e^{- 2D\gamma_p^0(t-t') - 2D k p^2 (\Lambda(t)-\Lambda(t'))} dt'.
\eeq
In the long-time limit, we expect the system to reach equilibrium.  Accordingly, we seek a solution of the form $\Lambda(t) = \lambda_0 t$, {\em i.e.}, 
constant $\lambda_0(t)$. $\chi_q(t,t')$ then depends only on the time difference $(t-t')$, and we are free to Fourier transform. The right hand side of the 
PIDE can be viewed as the Fourier transform of $\tilde \chi_q^2(t,0) \Theta(t)$ evaluated at zero frequency, which leads us immediately 
to Eq.~\ref{eq: delta ell equilibrium}.

As expected, this reproduces the equilibrium mean-field theory equation of Ref.~\cite{kernes2020equilibrium}. While, the sum can be 
performed in closed form, we approximate the summation by an integration in order to understand its $k$-dependence. 
Since deviations in $\lambda_0$ from the spring-free result occur at larger values of $k$, the distinction between the 
summation and integration is immaterial. In terms of the $k$-independent change in projected length $\lambda_0^\text{free}$ 
(found by setting $k=0$ in Eq.~\ref{eq: delta ell equilibrium}), we find the equation
\beq
\frac{\lambda_0}{\lambda_0^\text{free}} = \left[ 1 + \frac{k \lambda_0^\text{free}}{\tau} \left(\frac{\lambda_0}{\lambda_0^\text{free}}\right)\right]^{-1/2}.
\eeq
The most interesting result is found at high $k$, where the solution to this equation demands $\lambda_0/\lambda_0^\text{free} 
\sim k^{-1/3}$. Consequently, the effective tension $k \lambda_0 \sim k^{2/3}$. The transition occurs when $ k \lambda_0/\tau \approx 1$. 
These results are confirmed by Fig.~\ref{fig: deltaPhiK}.

We now consider the short-time limit, where the behavior is dependent on the initial condition. We treat the case where 
$u_p(0)$ is averaged over the $k\neq 0$ equilibrium ensemble, and at $t=0$ a small, additional time-dependent 
tension
\beq
\tau(t) = \tau + \delta f(t) \Theta(t),
\eeq
is applied. For reference, in equilibrium
\beq
\langle u_p^2(0) \rangle_\text{eq} = \frac{2k_\text{B} T/\ell}{\kappa p^2(p^2 + \tau + k \lambda_0)},
\eeq
which can be inferred from the long-time MFT solution. $\delta f(t)$ has magnitude $f$, and is superimposed on 
top of a prestress $\tau$. In analogy with defining the time-integrated projected length, we find it useful to introduce the time-integrated 
applied tension 
\beq
\label{eq: big F def}
\delta F(t) = \int_0^t \delta f(t') dt'.
\eeq 

Upon turning on the additional tension $\delta f(t)$, the projected length will change an amount 
$\delta \langle \Delta \ell(t) \rangle=\langle \Delta \ell(t) - \Delta \ell(0) \rangle$, and the integrated projected length will change by an 
amount $\Lambda(t) = \Lambda_0 + \delta \Lambda(t)$, where $\Lambda_0 = \lambda^\infty_0 t$ is the long-time constant solution. 
Comparing the two, we can identify
\beq
\label{eq: delta lambda ell}
\p_t \delta \Lambda = \delta \langle \Delta \ell \rangle.
\eeq
This relates $\delta \Lambda$ to the projected length response (which is not necessarily linear). 
Decomposing $\Lambda = \Lambda_0 + \delta \Lambda$, we redefine
\beq
\tilde \chi_q(t,t') = e^{-D q^2(\kappa q^2+\tau+ k \lambda_0)(t-t')}e^{-Dq^2 (\delta A(t) +\delta A(t')},
\eeq
where we have grouped the two perturbations $\delta \Lambda(t)$ and $\delta F(t)$ into a single function
\beq
\delta A(t) = k \delta \Lambda(t) + \delta F(t).
\eeq
This can similarly be accomplished by setting $\tau \to \tau + k\lambda_0 + \delta f(t)$ and 
replacing $\Lambda(t) \to \delta \Lambda(t)$ in Eq.~\ref{eq: chi app1}. Substituting and averaging over the initial condition yields the PIDE
\beq
\label{eq: delta lambda sum}
\frac{d \delta \Lambda}{dt} = \frac{k_\text{B}T}{2} \sum_q \left\{ \frac{\tilde \chi_q^2(t,0)-1}{\kappa q^2 + \tau + k \lambda_0}+ \frac{2q^2}{\xi_\perp} \int_0^t \tilde \chi_q^2(t,t') dt' \right\}.
\eeq

We are interested in the short-time solution to this equation. Since the projected length must be finite at $t=0$, this implies that at, 
short times, $\delta \Lambda(t) \sim t^{\eta}$ for some $\eta>1$. The prestress ensures that $\delta F(t)$ can be made small (by 
reducing the amplitude of applied tension) relative to $\tau$ at all values $q$, allowing one to expand $\delta F(t)$ in the exponential of 
$\tilde \chi_q(t,t')$ as $t \to0$~\cite{obermayer2009tension}. Consequently, the change $\delta \Lambda(t)$ will be small as well, since it vanishes at $f=0$. 
These considerations suggest that we can expand $\tilde \chi_q(t,t')$ in a power series about $\delta \Lambda(t)$ and $\delta F(t)$. Doing so, we find
\beq
\frac{d \delta \Lambda}{d t} = -\int_0^t  M(t-t') [k\delta \Lambda(t') + \delta F(t')] dt',
\eeq
where we have defined the kernel
\beq
M(t) = \sum_p \left[\frac{D p^2 \delta(t)}{\kappa p^2 + \tau +k \lambda_0} -2D^2 p^4 e^{-2Dp^2(\kappa p^2 + \tau + k \lambda_0)t} \right].
\eeq
This may be solved by Laplace transformation. The Laplace transform of the kernel is
\beq
\label{eq: M tilde}
\tilde M(z) = \sum_p \frac{zD p^2}{(\kappa p^2 + \tau+k \lambda_0)[z + 2D p^2(\kappa p^2 + \tau +k \lambda_0)]}.
\eeq
In terms of the dimensionless tension $\phi$, the shift $\Delta \phi$ defined in Eq.~\ref{eq: delta phi}, $\bar k$, and the dimensionless 
Laplace variable $\bar z = z \ell^4/D \kappa \pi^4$, we can equivalently write this as
\beq
\label{eq: M tilde phi}
\tilde M(\bar z) = \sum_{n=1}^\infty \frac{D n^2 \bar z/\kappa}{(n^2+\phi+\Delta \phi)[\bar z + 2 n^2(n^2 + \phi + \Delta \phi)]}
\eeq
Solving, the transformed change in projected length $\delta \Lambda(z)$ is 
\beq
\delta \Lambda(z) = -\frac{\tilde M(z)/z}{1 + k\tilde M(z)/z}\delta F(z).
\eeq
Since $F(z)$ is proportional to $f$, we may divide both sides by $f$, then use Eq.~\ref{eq: delta lambda ell} to obtain the 
Laplace transform of the projected length linear response
\beq
\label{eq: chi z}
\chi_{\Delta \ell}(z) = -z\frac{\tilde M(z)/z}{1 + k \tilde M(z)/z}\frac{\delta F(z)}{f}.
\eeq

\section{Polarization function calculation}
\label{app: polarization}
It is computationally easier to begin by working in the time domain. We decompose $\mathbf{M}^{-1}(t,t')$ in 
terms of its $k=0$ and $k \neq 0$ pieces via
\beq
\mathbf{M}^{-1}(t,t') =  \bm{\sigma} + D k \bm{\Pi}(t,t'),
\eeq
where $\bm{\sigma}$ represents the 2x2 block matrix with zeros along the diagonal, and identity matrices on the off diagonal. 
We call the additional contribution, $\bm{\Pi}(t,t')$,  the polarization matrix, in analogy to electron screening in metals~\cite{altland2010}.
It encodes fluctuation corrections, and is determined by the trace-log. Specifically, it is given by the second-order term in the 
Taylor expansion of $\Tr \ln (\mathbb{1} + D k p^2 \hat G \hat {\delta \lambda})$ about the small matrix
\beq
\hat{\delta \lambda}(t) = \left(
\begin{array}{cc}
 0 & \delta \lambda (t) \\
 \delta \lambda(t)  & -\delta \overline{ \lambda }(t) \\
\end{array}
\right).
\eeq 
$\hat G$ is the saddle-point, matrix-valued Green's function 
\beq
\hat G(t,t') =
\left(\begin{array}{cc} 
0 & \bar G^-_p(t-t') \\
\bar G^+_p(t-t') & \bar C_p(t-t')
\end{array} \right),
\eeq
with components given by the time representation of Eq.~\ref{eq: G MFT},
\beq
\bar G^\pm_p(t) = \Theta(\pm t) e^{\mp D \gamma_p t},
\eeq
and Eq.~\ref{eq: C MFT},
\beq
\bar C_p(t,t') = 2D \int d\tau G^+(t-\tau) G^-(\tau-t').
\eeq

The modified function $\gamma_p = \gamma_p^0 +k p^2 \lambda_0$, includes the saddle-point 
value $\lambda_0$.  The logarithm of matrices is defined via its Taylor series, whose quadratic term is 
$\frac{-1}{2}\Tr  \hat{G} \hat{\delta \lambda} \hat{G} \hat{\delta \lambda}$. The factor of $1/2$ can be factored out, per the definition of 
$\bm{\Pi}$. Products of the form $\hat G^\pm(t) \hat G^\pm(-t)$ have vanishing support due to the $\theta$ functions and are zero. 
Carrying out the matrix products, we find that $\bm{\Pi}(t,t')=\bm{\Pi}(t-t')$ is a function only of the time difference, with the result
\beq
\bm{\Pi}(t)=\sum_p p^4
\left(
\begin{array}{cc}
-C_p^2(t) & 2G_p^+(t) C_p(t) \\
 2G_p^-(t) C_p(t) & 0
 \end{array}
\right).
\eeq
Since each of the operators depend on only the time difference $t-t'$, we may Fourier transform to frequency space. 
Including the $\bm{\sigma}$ contribution we find the effective functional matrix 
\beq
\label{eq: M inverse}
\mathbf{M}_\omega^{-1} = \left( \begin{array}{cc} -\Pi^0_\omega & \Pi^+_\omega +1\\ \Pi^-_\omega + 1 & 0 \end{array} \right).
\eeq
The individual components are given by Eqs.~\ref{eq: pi plus} and ~\ref{eq: pi 0}.
We have chosen the $\pm$ notation to emphasize the similarity of $\Pi^\pm$ to Green's functions, and $\Pi^0$ to the spring-free 
correlator. Indeed, $\Pi^+ = (\Pi^-)^*$, and, as a consequence of the fluctuation-dissipation, the $\Pi$ functions obey the relationship
\beq
\label{eq: pi FDT}
\text{Im} \Pi^+_\omega = \frac{\omega}{2 k_\text{B} T} \Pi^0_\omega.
\eeq
We thus need only compute $\Pi^+$ to fully specify the polarization matrix. 

\section{transverse spring only}
\label{app: transverse spring}
For completeness, we report the solution of the problem for a purely transverse spring attached at the endpoint ({\em i.e.}, 
no longitudinal component). We follow the method of Ref.~\cite{kernes2020equilibrium} for dealing with inhomogeneous boundary 
conditions in Fourier space. In this section, primes refer to spatial derivatives.

The homogeneous boundary conditions are pinned, with zero torque at both endpoints: $u(x_S) = u''(x_S)=0$, and $x_S=0,\ell$. 
Wavenumbers are set to $p_n = n \pi/\ell$, for $n$ a positive integer. The transverse spring 
replaces the pinned boundary condition $u(\ell)=0$ with the new condition
\beq
\label{eq: transverse BC}
- \kappa u''''(\ell) + \tau u''(\ell) = -k_\perp u(\ell).
\eeq

In the bulk, we still have the linear Langevin equation
\beq
\p_t u+ D \kappa u'''' - D \tau u'' = \zeta(x,t),
\eeq
subject to the aforementioned boundary conditions. In order to implement the boundary condition, we add an 
additional force operator that is non-diagonal in wavenumber and regulated by a parameter $\epsilon$ that we take to 
zero at the end of the calculation~\cite{kernes2020equilibrium}. We write:
\beq
\label{eq: perp langevin}
 \left[\delta_{nm}\left(\p_t +D\kappa p_n^4 + D\tau p_n^2\right) + \frac{1}{4\epsilon} \psi_n \psi_m \right]u_m =  \delta_{nm} \xi_m,
\eeq
where we have defined the infinite dimensional vector
\beq
\psi_n =(-1)^{n}\left(\frac{\gamma_n }{p_n} + \frac{1}{2}k_\perp \sin 2n\pi \right).
\eeq
This is solved by the method of Green's functions. We replace $\zeta$ on the right side with a $\delta$-function in time and a 
Kronecker delta $\delta_{nk}$, and $u_m(t)$ by the Green's function $\chi_{mk}^\perp(t-t')$. The response is still given by a sum 
over sines
\beq
\label{eq: chi sum}
\chi^\perp(x,x';t) = \sum_{m,n=1}^\infty \chi_{mn}^\perp(t) \sin(p_n x) \sin (p_m x').
\eeq
Next, we Laplace transform the $\chi$ version of Eq.~\ref{eq: perp langevin}, take the inverse of the left side, and finally take the $\epsilon \to 0$
 limit to find
\beq
\label{eq: chi perp}
\chi_{mk}^\perp(s) = \chi_{mk}^\text{D}(s) + \chi_{mk}^\text{BC}(s),
\eeq
which has decomposed into a homogeneous part plus boundary term. The homogeneous part is
\beq
\label{eq: chi D}
\chi^\text{D}_{nm}(s)=\chi^\text{0}_{n}(s) \delta_{nm} = \frac{\delta_{nm}}{s + \gamma_n},
\eeq
where for this section we have defined
\beq
\gamma_n = D\kappa p_n^4 + D \tau p_n^2.
\eeq
The boundary term is given by
\beq
\chi_{mk}^\text{BC}(t)=\frac{-(\chi^0_{n} \psi_n)^2}{\sum_{n=1}^\infty \psi_n \chi^0_n \psi_n}.
\eeq. 
The numerator is
\beq
\text{numerator}= -\frac{(-1)^{n+m}}{p_n p_m} \left[\frac{\gamma_n \gamma_m}{(s + \gamma_n)(s + \gamma_m)}\right].
\eeq
The denominator is a divergent sum. It has two main pieces
\beq
\sum_{n=1}^\infty \left\{ \frac{\gamma_n^2}{p_n^2(s+\gamma_n)} + k_\perp \frac{\gamma_n \sin 2\pi n}{p_n(s+\gamma_n)} + \text{convergent} \right\}
\eeq
The third piece is a convergent sum proportional to $\sin 2\pi$, and can be safely set to zero. We rewrite the series by 
subtracting out the divergent pieces as
\beq
\sum_{n=1}^\infty \left\{ \frac{-s \gamma_n}{p_n^2(s+\gamma_n)} +\frac{\gamma_n}{p_n^2} + \frac{-s \sin 2\pi n}{p_n(s+\gamma_n)}+ \frac{k_\perp \sin 2\pi n}{p_n}\right\}
\eeq
The first and third are now convergent, so the third can immediately be set to zero. The second and fourth need regularization. 
These sums were computed previously~\cite{kernes2020equilibrium}, with the results $-\tau/2$ and $-k_\perp \ell/2$ respectively. 
We then have
\beq
\text{denominator} = -\frac{1}{2}(\tau+ k_\perp \ell) - \sum_{n=1}^\infty\frac{s (\kappa p_n^2 + \tau)}{s+ \kappa p_n^4 + \tau p_n^2}.
\eeq
Combining, we find the boundary response
\beq
\label{eq: chi BC}
\chi^\text{BC}_{mn}(s) =\frac{2(-1)^{n+m}\gamma_n \gamma_m}{p_n p_m(\tau+k_\perp \ell + F(s))(s + \gamma_n)(s + \gamma_m)},
\eeq
where
\beq
F(s) =2\sum_{n=1}^\infty\frac{s (\kappa p_n^2 + \tau)}{s+ \kappa p_n^4 + \tau p_n^2}.
\eeq
The sum of Eqs~\ref{eq: chi D} and~\ref{eq: chi BC} gives the final result for the Laplace-transformed response function for a 
purely transverse spring at the boundary. To obtain the space/time domain solution, one can numerically 
perform the inverse transform and sum over modes according to Eq.~\ref{eq: chi sum}.

\bibliography{references3}
\end{document}